\documentclass[peerreview]{IEEEtran}

\usepackage[square,numbers,sort&compress]{natbib}
\usepackage{hyperref} %
\usepackage%
{hyperref} %
\hypersetup{
    colorlinks,
    linkcolor={blue!80!black},%
    citecolor={green!50!black},
    urlcolor={blue!80!black}
}

\usepackage{amsthm}
\usepackage{amsmath}
\usepackage{mathrsfs}
\usepackage{amssymb} 
\usepackage{stackrel}
\usepackage{mathtools}
\usepackage{nicefrac}

\usepackage{physics}

\usepackage{verbatim}
\usepackage{enumerate}

\usepackage[T1]{fontenc}
\usepackage{bbm} 
\usepackage{bbding}
\usepackage{keystroke}
\usepackage{pifont}
\usepackage{relsize} %

\usepackage{graphicx}
\usepackage{xcolor}
\usepackage{tikz}
\usetikzlibrary{calc}
\usepackage{tabularx}
\usetikzlibrary{arrows,shapes}

\renewcommand{\trace}{\mathrm{Tr}}

\newcommand{\identity}{\mathbbm{1}}

\newcommand{\markovC}[1]{%
\begin{tikzpicture}[#1]%
\draw (0,0.3ex) -- (1ex,0.3ex);%
\draw (0.5ex,0.3ex) circle (0.2ex);
\draw[white] (0.2ex,0) -- (0.5ex,0);%
\end{tikzpicture}%
}
\newcommand{\Cbar}{\markovC{scale=2}}

\theoremstyle{remark}	\newtheorem{theorem}{Theorem}
\theoremstyle{remark}	\newtheorem{lemma}[theorem]{Lemma}
\theoremstyle{remark}	\newtheorem{corollary}[theorem]{Corollary}
\theoremstyle{remark}	
\theoremstyle{remark} \newtheorem{definition}{Definition}
\theoremstyle{remark} \newtheorem{remark}{Remark}
\theoremstyle{remark} \newtheorem{example}{Example}

\title{Quantum Relay Channels}

\author{
    \IEEEauthorblockN{Uzi Pereg\IEEEauthorrefmark{1}\IEEEauthorrefmark{2}
    } \\
		\vspace{0.25cm}
    \IEEEauthorblockA{\normalsize
		\IEEEauthorrefmark{1}Faculty of Electrical and Computer Engineering, Technion \\
		\IEEEauthorrefmark{2}Helen Diller Quantum Center, Technion \\
    Email: {\tt uzipereg@technion.ac.il
    }}
}

\date{\today}

\begin{document}

\maketitle

\begin{abstract}
Communication over a fully quantum relay channel is considered. %
We establish three bounds  based on different coding strategies, i.e., partial decode-forward, measure-forward, and assist-forward.
Using the partial decode-forward strategy,
the relay decodes part of the information, while the other part is decoded without the relay's help. The result by Savov et al. (2012)  for  a classical-quantum relay channel is obtained as a special case.
Based on our partial 
decode-forward bound,
 the capacity is determined for Hadamard relay channels. 
In the  measure-forward coding scheme, the relay performs a sequence of measurements and then sends a compressed representation of the measurement outcome to the destination receiver. The measure-forward strategy can be viewed as a generalization of the classical compress-forward bound.  %
At last, we consider quantum relay channels with orthogonal receiver components.
The assist-forward bound is based on a new approach, whereby the transmitter sends the message to the relay and simultaneously generates entanglement assistance between the relay and the destination receiver. Subsequently, the relay can transmit the message to the destination receiver with rate-limited entanglement assistance.
\end{abstract}


\maketitle

\begin{figure}[tb]
\begin{center}
\includegraphics[scale=0.5,trim={-3cm 11cm 0 11cm},clip]{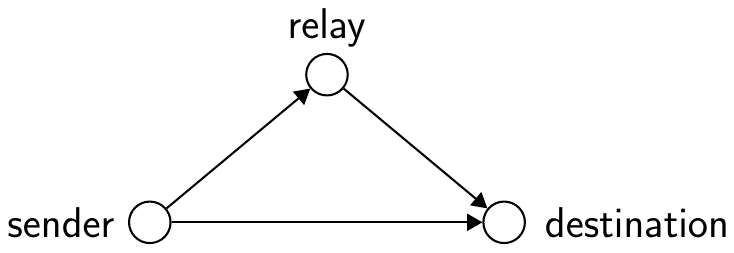} %
\end{center}
\caption{
A three-terminal relay network.
}
\label{Figure:Relay_Network}
\end{figure}

\begin{figure}[tb]
\begin{center}
\includegraphics[scale=0.6,trim={-3cm 11cm 0 11cm},clip]{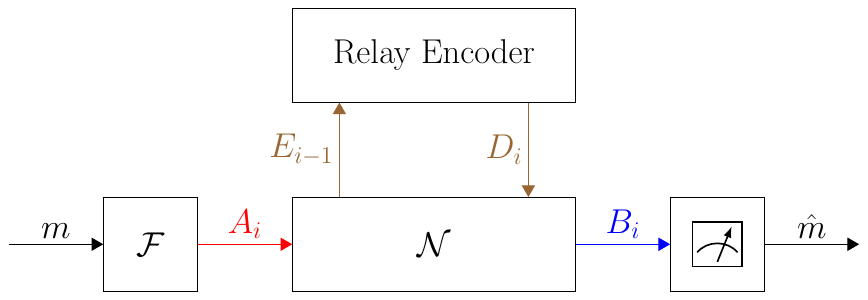} %
\end{center}
\caption{
 Coding for a fully quantum relay channel $\mathcal{N}_{A D \rightarrow B E}$.  The quantum systems of Alice,  Bob, and the relay are marked in red, blue,  and brown, respectively.  %
}
\label{Figure:Q_Relay_Coding}
\end{figure}

\section{Introduction}
Relaying plays a crucial role in enabling long-range communication. %
For example, in free-space optics systems, the transmission distance is limited by
 atmospheric conditions,
including absorption, scattering, and  turbulence \cite{VuPhamDangPham:20p}.
Attenuation in optical fibers poses a significant challenge as well. 
By dividing a communication link into 
two segments and placing a relay terminal between them,  one can rectify problems such as photon loss and operation defects \cite{AAUZHIA:23p}. %
Relaying is thus considered a promising and effective
solution \cite{XZSA:23p}.
Optimists view this solution as a key step toward quantum-enabled 6G communication \cite{%
BassoliFitzekBoche:23p}.
Furthermore, in the long-term vision of
an ad-hoc network of the quantum~Internet \cite{AEEHJLT:23p}, any node could act both as a transceiver of data and as a relay for other transmissions
\cite{GastparVetterli:02c}.

Cooperation in quantum communication networks has become a major focus of study in recent years, driven by advances in  experimental techniques and theoretical insights \cite{orieux2016recent,vanLoockAltBecherBensonBocheDeppe:20p,HGLLG:23p,luo2023recent,nemirovsky2024increasing}.
Entanglement is a valuable resource in network 
communication \cite{BFSDBFJ:20b}.
In the point-to-point setting, entanglement assistance between the transmitter and the receiver can significantly increase throughput
\cite{BennettShorSmolin:02p}, 
even if the resource is noisy \cite{ZhuangZhuShor:17p} or unreliable \cite{PeregDeppeBoche:23p}.
Entanglement assistance has recently been considered under the security requirements of secrecy \cite{QiSharmaWilde:18p,WuLongHayashi:22p,LedermanPereg:24c1} and covertness \cite{ZlotnickBashPereg:23c,ZlotnickBashPereg:23p,WanSuBloch:24c} as well.
In multi-user networks, entanglement between transmitters can also increase achievable rates for classical multiple-access channels
\cite{LeditzkyAlhejjiLevinSmith:20p,PeregDeppeBoche:23c,PeregDeppeBoche:23p2}, and yet entanglement between receivers  improves neither achievable rates \cite{PeregDeppeBoche:21p2}, nor error probabilities \cite{FawziFerme:24p}, for broadcast channels.
The three-terminal relay channel in Figure~\ref{Figure:Relay_Network} is a fundamental unit in user cooperation as well \cite{ChakrabartiSabharwalAazhang:06b}, and can also be used to generate entanglement between network nodes \cite{NatorPereg:24a1}.

In a multihop network, there are multiple rounds of communication, thus, synchronization is essential \cite{KramerGastparGupta:05p,Kramer:08n}. 
Suppose that the operation is governed by a central clock that ticks $n$ times.
Between the clock ticks $i-1$ and $i$, the sender and the relay transmit the channel inputs $A_i$ and $D_i$, respectively. See Figure~\ref{Figure:Q_Relay_Coding}.
Then, at the clock tick $i$,
the relay and destination receivers receive the channel outputs $E_i$ and $B_i$. This requires a small delay before reception to ensure causality.
The classical channel model was originally introduced by van der Meulen \cite{vanderMeulen:71p} as a building block for multihop networks \cite{PeregSteinberg:19p3}.

Savov et al. \cite{SavovWildeVu:12c,Savov:12z} considered  classical-quantum (c-q) relay channels and derived a partial decode-forward achievable rate %
for sending classical information. In the c-q case, both $A_i$ and $D_i$ are classical. Boche et al. \cite{BochCaiDeppe:15p} considered a c-q model of  two-phase bidirectional relaying. Communication with the help of environment measurement can be viewed as a quantum channel with a classical relay in the environment \cite{HaydenKing:04a}.
Considering this setting, Smolin et al. \cite{Smolin:05p} and Winter \cite{Winter:05a} determined the environment-assisted quantum capacity
and classical capacity, %
respectively. Furthermore, Ding et al. \cite{DingGharibyanHaydenWalter:20p} established the cutset, multihop, and coherent multihop bounds on the capacity of the c-q relay channel.
To the best of our knowledge, a fully quantum channel was not considered.

Network settings with causality aspects often require  block Markov coding \cite{ChoudhuriKimMitra:13p}, where the transmitter sends a sequence of blocks, and each block transmission encodes descriptions that are associated with the current and previous blocks. Quantum versions of block Markov coding were previously used for c-q relay channels \cite{SavovWildeVu:12c,Savov:12z}, communication with parameter estimation at the receiver \cite{Pereg:21p}, and quantum cribbing between transmitters \cite{PeregDeppeBoche:22p}. 

Sending quantum information is outside the scope of the present work, as we focus here on the classical task of sending a classical message.  Nonetheless,
we note that in order to distribute entanglement and send quantum information, the quantum relay would need to operate as a quantum repeater \cite{BriegelDurCiracZoller:98p}.  Pereg et al. \cite{PeregDeppeBoche:21p2}
addressed transmission of quantum information via a primitive relay channel, with a noiseless qubit pipe from the relay to the destination receiver, and provided an information-theoretic perspective on quantum repeaters. Other relay channel models for quantum repeaters can also be found in  \cite{GyongyosiImre:12c,JRXYLM:12p,GyongyosiImre:14c,Pirandola:16a,GhalaiiPirandola:20p}.

We consider the transmission of messages via a fully quantum relay channel. As opposed to previous work \cite{SavovWildeVu:12c,Savov:12z,BochCaiDeppe:15p,DingGharibyanHaydenWalter:20p}, the channel is fully quantum.
We establish three bounds  based on different coding strategies, i.e., partial decode-forward, measure-forward, and assist-forward.
Using the partial decode-forward strategy,
the relay decodes part of the information, while the other part is decoded without the relay's help.
Based on our partial 
decode-forward bound,
we determine the capacity for the special class of Hadamard relay channels. We also recover the result by Savov et al. \cite{SavovWildeVu:12c} for  the special case of a c-q relay channel.
In the  measure-forward coding scheme, the relay performs a sequence of measurements and then sends a compressed representation of the measurement outcome to the destination receiver. The measure-forward strategy can be viewed as a generalization of the classical compress-forward bound due to Cover and El Gamal \cite{CoverElGamal:79p}. 
At last, we consider quantum relay channels with orthogonal receiver components.
The assist-forward bound is based on a new approach, whereby the transmitter sends the message to the relay and simultaneously generates entanglement assistance between the relay and the destination receiver. Subsequently, the relay can transmit the message to the destination receiver with rate-limited entanglement assistance.
We demonstrate our results by computing a closed-form lower bound for a depolarizing relay channel. 

The paper is organized as follows. 
In Section~\ref{Section:Definitions}, we give preliminary definitions and present the channel model.
 Section~\ref{Section:Coding} provides the coding definitions for communication with strictly causal quantum relaying.
 Section~\ref{Sec:Main} presents our main results, including the partial decode-forward, measure-forward, and assist-forward bounds.
  In Section~\ref{Section:Examples}, we give  examples. 
 Section~\ref{Section:Summary} concludes with a summary and discussion. 
 The analysis is given in the appendix. Appendix~\ref{App:Tools} provides information-theoretic tools.
 In Appendices~\ref{Appendix:PDF} through \ref{Appendix:AF}, we derive the partial decode-forward, measure-forward, and assist-forward bounds, with Appendix~\ref{Appendix:Hadamard_Relay} addressing Hadamard relay channels.
 In Appendix~\ref{Appendix:Depolarizing}, we provide the analysis for the depolarizing relay channel.

\section{Definitions and Channel Model}%
\label{Section:Definitions}
\subsection{Notation, States, and Information Measures}
\label{subsec:notation}
 We use the following notation conventions. %
Script letters $\mathcal{X},\mathcal{Y},\mathcal{Z},...$ are used for finite sets.
Lowercase letters $x,y,z,\ldots$  represent constants and values of classical random variables, and uppercase letters $X,Y,Z,\ldots$ represent classical random variables.  
 The distribution of a  random variable $X$ is specified by a probability mass function (pmf) 
	$p_X(x)$ over a finite set $\mathcal{X}$. %
 We use $x^j=(x_1,x_{2},\ldots,x_j)$ to denote  a sequence of letters from $\mathcal{X}$. %
 A random sequence $X^n$ and its distribution $p_{X^n}(x^n)$ are defined accordingly. 

A quantum state is described by a density operator $\rho$ on the Hilbert space $\mathcal{H}$. The dimensions are assumed to be finite. %
 We denote the set of all such density operators by $\mathfrak{D}(\mathcal{H})$.
 The probability distribution of a measurement outcome can be described in terms of a positive operator-valued measure (POVM), i.e. a set  of positive semidefinite operators $\{ \Delta_j \}$, such that %
$\sum_j \Delta_j=\identity$, where $\identity$ is the identity operator. 
According to the Born rule, if the system is in state $\rho$, then the probability of the measurement outcome $j$ is given by $\Pr(j)=\trace(\Delta_j \rho)$.

Define the quantum entropy of the density operator $\rho$ as
$%
H(\rho) \equiv -\trace[ \rho\log(\rho) ]
$, %
which is the same as the Shannon entropy %
associated with the eigenvalues of $\rho$.
Consider the state of a pair of systems $A$ and $B$ on the tensor product $\mathcal{H}_A\otimes \mathcal{H}_B$ of the corresponding Hilbert spaces.
Given a bipartite state $\rho_{AB}$, %
define the quantum mutual information as
\begin{align}
I(A;B)_\rho=H(\rho_A)+H(\rho_B)-H(\rho_{AB}) \,. %
\end{align} 
Furthermore, conditional quantum entropy and mutual information are defined by
$H(A|B)_{\rho}=H(\rho_{AB})-H(\rho_B)$ and
$I(A;B|C)_{\rho}=H(A|C)_\rho+H(B|C)_\rho-H(AB|C)_\rho$, respectively.
The coherent information is then defined as
\begin{align}
I(A\rangle B)_\rho=-H(A|B)_\rho \,. %
\end{align}

\begin{figure}[tb]
\begin{center}
\includegraphics[scale=0.5,trim={-3cm 11cm 0 11cm},clip]{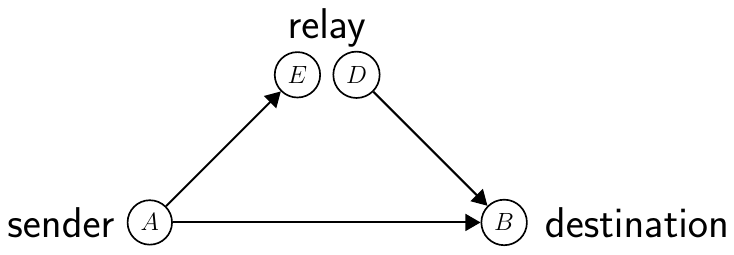} %
\end{center}
\caption{
A diagram of the quantum relay channel: The transmitters at the sender and the relay are labeled as $A$ and $D$, and the receivers at the destination and the relay as $B$ and $E$, respectively.
}
\label{Figure:Q_Relay_Network}
\end{figure}

\subsection{Channel Model}
\label{Subsection:Q_Relay_Channel}
We consider a fully quantum relay channel as a model for the three-terminal network in  Figure~\ref{Figure:Relay_Network}. 
A quantum relay channel is a completely-positive trace-preserving (CPTP) map, $\mathcal{N}_{A D \rightarrow B E}$, where
$A$, $D$, $B$, and $E$  are associated with the 
sender transmitter, the relay transmitter, the destination receiver, and the relay receiver, respectively. See Figure~\ref{Figure:Q_Relay_Network}.
We assume that the channel is memoryless. That is, if the systems $A^n=(A_1,\ldots,A_n)$ and $D^n=(D_1,\ldots,D_n)$ are sent through $n$ channel uses, then the input state $\rho_{ A^n D^n}$ undergoes the tensor product mapping
\begin{align}
\mathcal{N}_{ A^n D^n\rightarrow B^n E^n}\equiv  (\mathcal{N}_{ A D\rightarrow B E })^{\otimes n}
\,.
\label{Equation:Block}
\end{align}
The communication setting will be defined in Section~\ref{Section:Coding} such that the relay encodes in a strictly causal manner. That is, 
the relay transmits $D_i$ at time $i$, and only then receives $E_i$.

\subsection{Special Cases}%
\label{subsec:Hadamard}
We will also discuss the special class of degraded relay channels.
Intuitively, if a relay channel is degraded, then  the output of the destination receiver is a noisy version of that of the relay.
In practice, this is typically the case, since the relay is an intermediate station between the transmitter and the destination receiver. 

 Let $\mathcal{M}_{AD\to E}$  denote the marginal channel to the relay, and $\mathcal{L}_{AD\to B}$ to the destination receiver:
\begin{align}
\mathcal{M}_{AD\to E}&=
\trace_B \circ \mathcal{N}_{AD\to BE} \,,
\\
\mathcal{L}_{AD\to B}&=
\trace_E \circ \mathcal{N}_{AD\to BE} \,.
\end{align}

\begin{definition}[Degraded relay channel] %
\label{Definition:Degraded}
  A quantum relay channel $\mathcal{N}_{AD\rightarrow B E}$ is called \emph{degraded} if there exists a degrading channel $\mathcal{P}_{E\rightarrow BE}$ such that the marginal satisfies the following relation,
	\begin{align}
	\mathcal{N}_{AD\rightarrow B E}%
    =\mathcal{P}_{E\to BE}\circ\mathcal{M}_{AD\to E} \,.
	\end{align}
In this case, we say that Bob's channel %
is degraded with respect to the relay's channel, $\mathcal{M}_{AD\to E}$.
\end{definition}

We also introduce the definition of a Hadamard relay channel. In this case, the relay receives a classical observation $Y_1$, which can be interpreted as a measurement outcome. 
\begin{definition}
\label{Definition:Hadamard_Relay}
A Hadamard relay channel $\mathcal{N}^{\,\text{H}}_{AD\rightarrow B Y_1}$
is a quantum-quantum-quantum-classical  relay channel that is also degraded, i.e.,
	\begin{align}
	\mathcal{N}^{\,\text{H}}_{AD\rightarrow B Y_1} %
    =\mathcal{P}_{Y_1 \to BY_1}\circ\mathcal{M}_{AD\to Y_1} 
	\end{align}
where $Y_1$ is classical. 
\end{definition}
Intuitively, if a relay channel is degraded, then  the output state of the destination receiver is a noisy version of that of the relay. 
Additional degraded classes are reviewed in the discussion section (see Subsection~\ref{Discussion:Degraded}).
For a Hadamard relay channel $\mathcal{N}^{\,\text{H}}_{AD\to B Y_1}$, the relay can be viewed as a measure-and-prepare device. The marginal channel $\mathcal{M}_{AD\rightarrow Y_1}$ acts as a measurement device, while the degrading channel $\mathcal{P}_{Y_1\rightarrow BY_1}$ corresponds to state preparation.

In this case, Bob's channel $\mathcal{L}_{AD\rightarrow B}$ is said to be entanglement-breaking \cite{Shor:02p}.
The 
map $\mathcal{N}^{\,\text{H}}_{AD\rightarrow B Y_1}$ can also be viewed as a Hadamard broadcast channel \cite{WangDasWilde:17p,PeregDeppeBoche:21p2}, where Bob and the relay are interpreted as two receivers.

\section{Coding}
\label{Section:Coding}
We consider a fully quantum relay channel $\mathcal{N}_{AD\to BE}$. Below, we define a code for the transmission of classical information via the quantum relay channel. 
The relay serves the purpose of assisting the transmission of information from Alice to Bob.  

\begin{definition} 
\label{def:ClcapacityE}
A $(2^{nR},n)$ classical code for the 
quantum relay channel 
$\mathcal{N}_{AD\to BE}$ 
consists of the following: 
\begin{itemize}
\item 
A message set  $[1:2^{nR}]$, where $2^{nR}$ is assumed to be an integer,  
\item
an encoding map $\mathcal{F}_{M \rightarrow A^n}:[1:2^{nR}]\to \mathfrak{D}(\mathcal{H}_A^{\otimes n})$,

\item
a sequence of strictly-causal relay encoding maps 
$\Gamma^{(i)}_{E_{i-1} \bar{E}^{i-2} \rightarrow  \bar{E}^{i-1} D_i}: \mathfrak{D}(\mathcal{H}_E^{\otimes (i-1)})\to  \mathfrak{D}(\mathcal{H}_E^{\otimes (i-1)}\otimes \mathcal{H}_D)$, for 
$i\in [1:n]$,

\item	
a decoding POVM   $\{\Delta_m\}  $ on 
$\mathcal{H}_B^{\otimes n}$, where the measurement outcome $m$ is an index in 	$[1:2^{nR}]$.
\end{itemize}
We denote the code by $(\mathcal{F},\Gamma,\Delta)$.
\end{definition}

The communication scheme is depicted in Figure~\ref{Figure:Q_Relay_Coding}.  
The systems $A^n$ and $D^n$ represent the transmissions by the sender and the relay,    respectively.
Whereas, $E^n$ and $B^n$ are the received outputs at the relay and the destination receiver, respectively.
Alice selects a uniform message $m\in [1:2^{nR}]$ that is intended for the destination receiver, Bob.
She encodes the message by applying the encoding map $\mathcal{F}_{  M \to A^n}$,
 and transmits the systems $A^n$ over $n$ channel uses.

At time $i$, the relay  applies the encoding map 
$\Gamma^{(i)}_{E_{i-1} \bar{E}^{i-2} \rightarrow  \bar{E}^{i-1} D_i}$ to $E_{i-1},\bar{E}^{i-2}$, and then receives $E_i$, for $i\in [1:n]$.  
The system $\bar{E}^{i-1}$ can be viewed as a ``leftover" of the encoding operation at time 
$i$. 
Specifically, 
at time $i=1$, we have
\begin{align}
\rho_{A^n D_1}^{(m)}&=
\mathcal{F}_{M\to A^n}(m)
\otimes
\Gamma^{(1)}_{E_0\to D_1}(1) \,,
\intertext{
where $E_0$ is a degenerate system of dimension $1$. 
We now split $A^n$ into
$A_1$ and $A_2^{n}\equiv (A_i)_{i\geq 2}$. 
Then, the relay receives $E_1$ in the state}
\rho_{ B_1 E_1 A_2^n}^{(m)} &=
\left(\mathcal{N}_{AD\to BE}\otimes \mathrm{id}_{A_2^n}\right)(\rho_{A_1 D_1 A_2^n}^{(m)}) 
\,.
\end{align}

At time $i=2$, the relay encodes $D_2$ by
\begin{align}
\rho_{\bar{E}_1 D_2 A_2^n  B_1 }^{(m)}&=
\left(
\Gamma^{(2)}_{E_{1}  \rightarrow  \bar{E}_1 D_2}
\otimes
\mathrm{id}_{A_2^n  B_1}
\right)
(\rho_{E_1 A_2^n   B_1   }^{(m)})
\,,
\intertext{
where $\bar{E}_1$ is a leftover that can be used in subsequent steps.
The relay receives $E_2$ in the state}
\rho_{ B_2  E_2 A_3^n B_1 \bar{E}_1  }^{(m)} &=
(\mathcal{N}_{AD\to BE}\otimes \mathrm{id}_{ A_3^n B_1 \bar{E}_1 })(\rho_{A_2 D_2 A_3^n B_1 \bar{E}_1 }^{(m)}) 
\,.
\end{align}

This continues in the same manner. 
At time $i=n$, the relay encodes by
\begin{align}
\rho_{\bar{E}^{n-1} D_n A_n  B^{n-1} }^{(m)}&=
\left(
\Gamma^{(n)}_{E_{n-1}
\bar{E}^{n-2}
\to \bar{E}^{n-1} D_n}
\otimes
\mathrm{id}_{A_n  B^{n-1}}
\right)(\rho_{E_{n-1}  \bar{E}^{n-2} A_n
B^{n-1}   }^{(m)})
\,,
\intertext{and then,}
\rho_{ B^n E_n \bar{E}^{n-1} }^{(m)} &=
(\mathcal{N}_{AD\to BE}\otimes \mathrm{id}_{  B^{n-1} \bar{E}^{n-1}})(\rho_{A_n D_n  B^{n-1} \bar{E}^{n-1} }^{(m)}) 
\,.
\end{align}
Thus, the output state at the destination receiver is the reduced state, 
$
\rho_{B^n }^{(m)}=
\trace_{E_n \bar{E}^{n-1}}\left(\rho_{ B^n E_n \bar{E}^{n-1} }^{(m)}\right) 
$. 
 
Bob receives the channel output system $B^n$, performs the measurement
$\{\Delta_m\}  $, and obtains  an estimate  $\hat{m}\in [1:2^{nR}]$ of Alice's message, as the measurement outcome.
The average  probability of error of the code  $(\mathcal{F},\Gamma,\Delta)$  
is given by 
\begin{align}
P_{e}^{(n)}(\mathcal{F},\Gamma,\Delta)&= 
1-\frac{1}{2^{nR}}\sum_{m=1}^{2^{nR}}\trace(  \Delta_{m}  \rho^{(m)}_{ B^n}) \,.
\end{align}

A $(2^{nR},n,\varepsilon)$  code for the quantum relay channel satisfies 
$
P_{e}^{(n)}(\mathcal{F},\Gamma,\Delta)\leq\varepsilon $.  
A rate  $R\geq 0$ is called achievable   if for every $\varepsilon,\delta>0$ and sufficiently large $n$, there exists a 
$(2^{n(R-\delta)},n,\varepsilon)$ code. 
Equivalently,   $R$ is  achievable   if there exists a sequence of codes at rate $R-\delta$ such that the average probability of error tends to zero as $n\to \infty$, where $\delta>0$ is arbitrarily small. 

 The  capacity of the quantum relay channel 
 $C(\mathcal{N})$ 
is defined as the supremum of achievable rates. 

\begin{remark}
\label{Remark:Relay_Decoding}
The relay is only required to assist the transmission of information to Bob, and does not necessarily decode any information. In particular, we will later introduce a measure-forward strategy where the relay does not decode information at all.
See Section~\ref{Subsection:MF} below.
\end{remark}

\begin{remark}
\label{Remark:Leftover}
The code definition above for the quantum setting is somewhat more involved than in previous work \cite{SavovWildeVu:12c,Savov:12z,BochCaiDeppe:15p,DingGharibyanHaydenWalter:20p}, since we need to account for the post-operation state at the relay in each time instance. That is, the relay encoding at time $i$ affects the state of $E_{i-1}$, as well as the previous outputs. 
As mentioned above,
the system $\bar{E}^{i-1}$ can be viewed as a ``leftover" of the relay encoding at time $i$. This leftover can be used in subsequent steps, at time 
$j>i$. For example, if the relay performs a sequence of measurements, then $\Gamma^{(i)}$ are quantum instruments such that $\bar{E}^{i-1}$ is the post-measurement system, while $D_i$ stores the measurement outcome.
In the classical setting, $\bar{E}^{i-1}$ is simply a copy of the previously received observations at the relay.
\end{remark}

\section{Main Results}
\label{Sec:Main}

Now, we give our results on the quantum relay channel
$\mathcal{N}_{AD\to BE}$. We will establish achievable rates that are based on different coding strategies.
We will also discuss special cases in which those strategies achieve the capacity of the quantum relay channel.

\subsection{Partial Decode-Forward Strategy}
\label{Subsection:PDF}
In the partial decode-forward strategy, the relay decodes part of the information, while the other part is decoded without the relay's help.
Our derivation can be viewed as a quantum version of  the classical coding strategy, under the same name \cite{CoverElGamal:79p}. 
Let
\begin{align}
\mathsf{R}_{\text{PD-F}}(\mathcal{N})\equiv \max_{  p_{U X_0 X_1} \,,\; \theta^{x_0}_{A}\otimes \zeta^{x_1}_{D}  } 
\min\bigg\{&
I(X_0 X_1;B)_{\omega}
\,,\;
I(U;E|X_1)_{\omega}
+I(X_0;B| X_1 U)_{\omega}
\bigg\}
\label{eq:low_PDF}
\end{align}
where the maximum is over the set of all probability distributions $p_{U X_0 X_1}(u,x_0,x_1)$ and product state collections $\{ \theta^{x_0}_{A}\otimes \zeta^{x_1}_D  \}$, with
\begin{align}
\label{eq:StateMaxCl}
&\omega_{U X_0 X_1 B E}=
\sum_{(u,x_0,x_1)\in\mathcal{U}\times \mathcal{X}_0\times \mathcal{X}_1}  p_{U X_0 X_1}(u,x_0,x_1) \ketbra{u}\otimes  \ketbra{x_0} \otimes \ketbra{x_1} \otimes \mathcal{N}_{AD\rightarrow BE}( \theta^{x_0}_{A} \otimes \zeta^{x_1}_D )
\,.
\end{align}

\begin{remark}
\label{Remark:External_U}
The random variable $U$ represents the information that is decoded by the relay, which gives rise to the mutual information term  $I(U;E|X_1)$.
The other terms are associated with the decoding at the destination receiver.
\end{remark}
\begin{remark}
\label{Remark:Direct_Full_DF}
Taking $U$ to be null, we obtain the direct-transmission lower bound:
\begin{align}
\mathsf{R}_{\text{PD-F}}(\mathcal{N})\geq  \max_{  p_{X_0} \,,\; \theta^{x_0}_{A}\otimes \ketbra{\zeta_{D}}  } 
I(X_0 ;B)_{\omega}
 \,.
\label{eq:low_Direct}
\end{align}
As expected, the direct-transmission lower bound is  the Holevo information \cite{Holevo:98p}.
To achieve this bound, the relay does not need to decode anything.
On the other hand, by taking $U=X_0$, we obtain 
\begin{align}
\mathsf{R}_{\text{PD-F}}(\mathcal{N})\geq \max_{  p_{X_0 X_1} \,,\; \theta^{x_0}_{A}\otimes \zeta^{x_1}_{D}  } 
\min\bigg\{&
I(X_0 X_1;B)_{\omega}
\,,\;
I(X_0;E|X_1)_{\omega}
\bigg\}
 \,.
\label{eq:low_Full_DF}
\end{align}
This bound is achieved through a \emph{full decode-forward} coding strategy, where the relay decodes the entire information.
As can be seen in the formula, this induces a bottleneck behavior.
\end{remark}

 Our first capacity result is given in the theorem below.
\begin{theorem}
\label{theo:PDF}
The capacity  of the  quantum relay channel 
$\mathcal{N}_{AD\to BE}$
satisfies
\begin{align}
C(\mathcal{N})\geq
\mathsf{R}_{\text{PD-F}}(\mathcal{N}) \,.
\end{align}
\end{theorem}
The proof of Theorem~\ref{theo:PDF} is given in Appendix~\ref{Appendix:PDF}.
Next, we consider three special classes.

\subsubsection{Classical-quantum channel}
As a special case, we recover the result by Savov et al. \cite{SavovWildeVu:12c}, for a classical-quantum channel.
Here, the transmissions $A$ and $D$ are replaced by  classical transmissions, $X$ and $X_1$, respectively. 
\begin{corollary}[{see \cite[Th. 1]{SavovWildeVu:12c}}]
\label{Corollary:CQ_Channel}
For a classical-quantum relay channel $\mathcal{N}_{X X_1\to B E}$,
\begin{align}
C(\mathcal{N})\geq 
\max_{  p_{U X X_1}   } 
\min\bigg\{&
I(X X_1;B)_{\omega}
\,,\;
I(U;E|X_1)_{\omega}
+I(X;B| X_1 U)_{\omega}
\bigg\} 
\label{eq:low_PDF_cq}
\end{align}
 where the maximum is over the set of all probability distributions $p_{U X X_1}(u,x,x_1)$, with
\begin{align}
\label{eq:StateMax_PDF_cq}
&\omega_{U X X_1 B E}=
\sum_{(u,x,x_1)\in\mathcal{U}\times \mathcal{X}\times \mathcal{X}_1}  p_{U X X_1}(u,x,x_1) \ketbra{u}\otimes  \ketbra{x} \otimes \ketbra{x_1} \otimes \mathcal{N}_{XX_1\rightarrow BE}(x,x_1 )
\,.
\end{align}
\end{corollary}

\begin{remark}
\label{Remark:Anti_Degradable}
In particular,
one may consider the trivial case of an
\emph{anti}-degradable channel.
Suppose that there exist reversely-degrading channels $\{\bar{\mathcal{P}}^{(x_1)}_{B\to E}\}$ %
such that 
\begin{align}
\mathcal{M}_{X X_1\to E}(x,x_1)=
\left(\bar{\mathcal{P}}^{(x_1)}_{B\to E} \circ \mathcal{L}_{X X_1\to B}\right) (x,x_1)
\end{align}
for all input pairs $(x,x_1)$.
Intuitively, the direct channel to the destination receiver is better than the channel to the relay, and thus, the relay is useless. As expected, the capacity is
\begin{align}
C(\mathcal{N})=
\max_{x_1\in\mathcal{X}_1}
\max_{  p_{X}   } 
I(X ;B|X_1=x_1)_{\omega}
\,.
\label{eq:low_PDF_cq_anti_degraded}
\end{align}
Achievability follows by taking 
$U$ to be null in \eqref{eq:low_PDF_cq} (see Remark~\ref{Remark:Direct_Full_DF}).
The converse proof is straightforward as well, based on standard arguments.
\end{remark}

\subsubsection{Hadamard relay channel}
Now, suppose the encoder and the relay both have quantum transmissions, $A$ and $D$, respectively, and the decoder receives a quantum system $B$. However, the relay receives a classical observation $Y_1$, which can be interpreted as a measurement outcome. Recall from Subsection~\ref{subsec:Hadamard}, that a Hadamard relay channel is also degraded. In this case, the relay can be interpreted as a measure-and-prepare device.   
\begin{theorem}
\label{Theorem:Hadamard_Relay}
The capacity of a Hadamard relay channel $\mathcal{N}_{AD\to B Y_1}$, as in Definition~\ref{Definition:Hadamard_Relay}, is given by
\begin{align}
C(\mathcal{N})= 
\max_{  p_{X_0 X_1} \,,\; \theta^{x_0}_{A}\otimes \zeta^{x_1}_{D}  } 
\min\bigg\{&
I(X_0 X_1;B)_{\omega}
\,,\;
I(X_0;Y_1|X_1)_{\omega}
\bigg\}
\label{eq:low_PDF_Hadamard}
\end{align}
where the maximum is 
as in \eqref{eq:low_PDF}.
\end{theorem}
The proof for Theorem~\ref{Theorem:Hadamard_Relay} is given in Appendix~\ref{Appendix:Hadamard_Relay}.
Essentially, Theorem~\ref{Theorem:Hadamard_Relay} says that
the 
\emph{full decode-forward} strategy is optimal
(see Remark~\ref{Remark:Direct_Full_DF}).
That is, the capacity is achieved when the relay decodes the entire information.
This is intuitive since for a Hadamard relay channel, the relay's channel is better than Bob's channel.

\subsubsection{Stinespring Dilation}
Suppose that the quantum relay channel 
$\mathcal{N}_{AD\to BE}$ is a Stinespring dilation of the channel $\mathcal{L}_{AD\to B}$ to Bob, and thus the channel $\mathcal{M}_{AD\to E}$ to the relay is a complementary channel to $\mathcal{L}_{AD\to B}$. 
This means that the quantum relay channel can be represented by an isometry $V:\mathcal{H}_A\to \mathcal{H}_B\otimes \mathcal{H}_E$.
The relay's system $E$ can then be viewed as Bob's environment. 
Furthermore, the full decode-forward bound yields the bound
\begin{align}
C(\mathcal{N})&\geq 
\max_{  p_{X_0 } \,,\; \ket{\theta^{x_0}_{A}}\otimes \ket{\zeta_{D}}  } 
\min\bigg\{
H(B)_{\omega}
\,,\;
H(E)_{\omega}
\bigg\}%
\end{align}
which resembles the environment-assisted distillation rate in \cite[Th. 1]{Smolin:05p} (see \cite{Winter:05a} as well).

\subsection{Measure-Forward Strategy}
\label{Subsection:MF}
We introduce a new coding scheme, where the relay performs a sequence of measurements, and then sends a compressed representation of the measurement outcome.
The relay does not decode any information in this coding scheme.
This can be viewed as a quantum variation of  the classical 
``compress-forward coding" approach \cite{CoverElGamal:79p}. 
Let
\begin{align}
\mathsf{R}_{\text{M-F}}(\mathcal{N})\equiv \max 
I(X_0 ;Z_1 B|X_1)_{\omega}
\label{eq:low_MF}
\end{align}
where the maximum is over the set of all product ensembles
$\{ p_{X_0},\theta^{x_0}_{A}\}$ 
$\otimes$
$\{ p_{X_1},\zeta^{x_1}_{D}\}$,
POVM collections
$\mathcal{G}_{E\to Y_1} =
\{ \Gamma_{y_1} \} $, and classical channels $p_{Z_1|X_1 Y_1}$ that satisfy
\begin{align}
I(Z_1;Y_1|X_1 B)_\omega \leq I(X_1;B)_\omega
\label{Equation:I_Z1_Y1_B}
\end{align}
for
\begin{align}
\omega^{(x_0,x_1)}_{BE}&= 
\mathcal{N}_{AD\rightarrow BE}( \theta^{x_0}_{A} \otimes \zeta^{x_1}_D )
\,,
\label{eq:MF_Output_1}
\\
\omega^{(x_0,x_1)}_{BY_1 Z_1}&=
\sum_{y_1\in  \mathcal{Y}_1}
\bigg[
\trace_E\left\{
\left( \identity\otimes \Gamma_{y_1} \right)\omega^{(x_0,x_1)}_{BE}
\right\}\otimes \ketbra{y_1}
\otimes
\label{eq:MF_Measurement_1}
\nonumber\\
& 
\sum_{z_1\in\mathcal{Z}_1}
p_{Z_1|X_1 Y_1}(z_1|x_1,y_1)
\ketbra{z_1} \bigg]
\,,
\\
\omega_{X_0 X_1 B Y_1 Z_1}&=
\sum_{(x_0,x_1)\in\mathcal{X}_0\times \mathcal{X}_1} p_{X_0}(x_0)p_{X_1}(x_1) \ketbra{x_0}\otimes
\ketbra{x_1}\otimes
\omega^{(x_0,x_1)}_{BY_1 Z_1}
\,.
\end{align}

\begin{remark}
\label{Remark:MF}
Intuitively,
the formula for $\mathsf{R}_{\text{M-F}}(\mathcal{N})$ above can be interpreted as follows. 
Given $x_0$ and $x_1$, the encoder and the relay prepare $\theta^{x_0}_{A}$ and $\zeta^{x_1}_D$, respectively. This results in the output state 
$\omega_{BE}^{(x_0,x_1)}$ in \eqref{eq:MF_Output_1}. The relay receives $E$ and performs a measurement, which yields $Y_1$ as an outcome. See \eqref{eq:MF_Measurement_1}.
The measurement outcome is then compressed and encoded by $Z_1$.

Roughly speaking, we interpret   $n I(Z_1;Y_1|X_1 B)_\omega$ as the number of information bits that are obtained from the measurement compression operation, while  the number of information bits that the relay can send through the channel to Bob is below $nI(X_1;B)_\rho$. This limitation is reflected through the maximization constraint in \eqref{Equation:I_Z1_Y1_B}.
\end{remark}

 Our measure-forward result is given  below.
\begin{theorem}
\label{theo:MF}
The capacity  of the  quantum relay channel 
$\mathcal{N}_{AD\to BE}$
satisfies
\begin{align}
C(\mathcal{N})\geq
\mathsf{R}_{\text{M-F}}(\mathcal{N}) \,.
\end{align}
\end{theorem}
The proof of Theorem~\ref{theo:MF} is given in Appendix~\ref{Appendix:MF}.

\subsection{Assist-Forward Strategy}
\label{Subsection:AF}
We introduce a new approach, whereby Alice simultaneously sends the message to the relay and generates entanglement assistance between the relay and Bob. Subsequently, the relay communicates the message to Bob using  rate-limited entanglement assistance.
We consider a quantum relay channel with orthogonal receiver components (ORC), where Bob's output has two components, i.e., $B=(B_1,B_2)$, and
the quantum relay channel $\mathcal{N}_{AD\to B_1 B_2 E}$ has the following form:
\begin{align}
\mathcal{N}_{AD\to B_1 B_2 E}=
\mathcal{M}_{A\to B_1 E}\otimes \mathcal{P}_{D\to B_2}
\,.
\label{Equation:Ortho_Receiver}
\end{align}
\begin{remark}
\label{Remark:ORC}
In words,
the quantum relay channel $\mathcal{N}_{AD\to B_1 B_2 E}$ above is a product of two channels, 
 a broadcast channel $\mathcal{M}_{A\to B_1 E}$ from the sender to the relay and the receiver, and a direct channel $\mathcal{P}_{D\to B_2}$ from the relay to the receiver. 
 The relay channel is referred to as having ORC due to this decoupling. 
\end{remark}

Consider the broadcast channel $\mathcal{M}_{A\to B_1 E}$ from the sender to the relay and the receiver. 
Given an ensemble $\{ p_{X_1},\theta^{x_1}_{G_0 G_1 A}\}$, let
\begin{align}
Q(\mathcal{M},\theta)&\equiv  \min \left\{
I(G_0 \rangle B_1 X_1 )_\theta
\,,\;
I(G_1 \rangle E X_1 )_\theta 
\right\}
\,.
\label{Equantion:Assistance_Capacity}
\end{align}
Moving to the quantum relay channel $\mathcal{N}_{AD\to B_1 B_2 E}$ with ORC, define
\begin{align}
\mathsf{R}_{\text{A-F}}(\mathcal{N})&\equiv
\max_{\theta\otimes \zeta} 
\min\left\{
I(X_1;E)_\theta \,,\;
I(X_2 G_2;B_2)_\zeta \,,\;
 I(X_2;B_2)_\zeta+ I(G_2\rangle B_2 X_2)_\zeta +Q(\mathcal{M},\theta)
\right\} \,.
\label{Equation:R_AF}
\end{align}
The maximum is %
over the set of all product ensembles
$\{ p_{X_1},\theta^{x_1}_{G_0 G_1 A}\}$ $\otimes$
$\{ p_{X_2},\zeta^{x_2}_{G_2 D}\}$, 
with
\begin{align}
\theta_{X_1 G_0 G_1 B_1 E}&=
\sum_{x_1\in\mathcal{X}_1} p_{X_1}(x_1) \ketbra{x_1}\otimes 
(\mathrm{id}_{G_0 G_1}\otimes \mathcal{M}_{A\rightarrow B_1 E})( \theta^{x_1}_{G_0 G_1 A }) \,,
\intertext{and}
\zeta_{X_2 G_3 B_2}&=
\sum_{x_2\in\mathcal{X}_2} p_{X_2}(x_2) \ketbra{x_2}\otimes 
(\mathrm{id}_{G_2}\otimes \mathcal{P}_{D\rightarrow B_2})( \zeta^{x_2}_{G_2 D })
\,.
\label{eq:EAF_Output}
\end{align}

\begin{remark}
\label{Remark:AF}
The formulas for $Q(\mathcal{M},\theta)$ and $\mathsf{R}_{\text{A-F}}(\mathcal{N})$ above are interpreted as follows. 
The input to the broadcast channel $\mathcal{M}_{A\to B_1 E}$ is chosen from the ensemble  $\{\theta^{x_1}\}$. 
Alice can send the message to the relay in this manner at rate %
$R<I(X_1;E)_\theta$.
Alice also generates entangled pairs, $G_0^n$ and $G_1^n$.
She simultaneously uses the broadcast channel $\mathcal{M}_{A\to B_1 E}$ to distribute the entanglement between Bob and the relay, respectively.
The entanglement rate is roughly $Q(\mathcal{M},\theta)$, as in \eqref{Equantion:Assistance_Capacity}.
In the subsequent block, the relay and Bob perform an entanglement-assisted communication protocol, using the entanglement resources that were generated in the previous block. 
This requires %
both 
$R<I(X_2 G_2;B_2)_\zeta$ and $R< I(X_2;B_2)_\zeta+ I(G_2\rangle B_2 X_2)_\zeta +Q(\mathcal{M},\theta)$, based on Shor's result \cite{Shor:04p} on rate-limited entanglement assistance.
\end{remark}

 Our assist-forward result is given  below.
\begin{theorem}
\label{theo:AF}
Consider a quantum relay channel %
with ORC, $B_1$ and $B_2$,
as in \eqref{Equation:Ortho_Receiver}.
The capacity  of such a channel %
satisfies
\begin{align}
C(\mathcal{N})\geq
\mathsf{R}_{\text{A-F}}(\mathcal{N}) \,.
\end{align}
\end{theorem}
The proof is given in Appendix~\ref{Appendix:AF}
combining  four fundamental techniques in quantum information theory:
\begin{enumerate}
\item
block Markov coding,
\item
constant-composition coding,
\item
rate-limited entanglement assistance, and 
\item
broadcast subspace transmission, 
\end{enumerate}
due to Cover et al. \cite{CoverElGamal:79p}, 
Winter \cite{Winter:99p}, Shor \cite{Shor:04p}, and Dupuis et al. \cite{DupuisHaydenLi:10p}, respectively. 

\begin{remark}
\label{Remark:EA_1}
Intuitively, our result suggests that even if the relay channel is particularly noisy, hence the entanglement rate
$Q(\mathcal{M},\theta)$ is small, the relay can still be useful. 
\end{remark}

\begin{remark}
\label{Remark:EA_2}
Earlier, we highlighted
that if a classical-quantum relay channel is anti-degraded, then    direct transmission is optimal. See Remark~\ref{Remark:Anti_Degradable}. Intuitively, the relay is useless in this case. 
This may appear to contradict Remark~\ref{Remark:EA_1}.
However, the conclusion in Remark~\ref{Remark:EA_1} only 
applies under the following conditions: (i) Alice can generate entanglement between the relay and the receiver. (ii) The
direct relay-receiver channel $\mathcal{P}_{D\to B}$   can benefit from entanglement assistance.
In contrast, both conditions do not apply to a classical-quantum  channel. %
\end{remark}

\section{Examples}
\label{Section:Examples}
\begin{figure}[tb]
\begin{center}
\includegraphics[scale=0.8,trim={-3cm 11cm 0 11cm},clip]{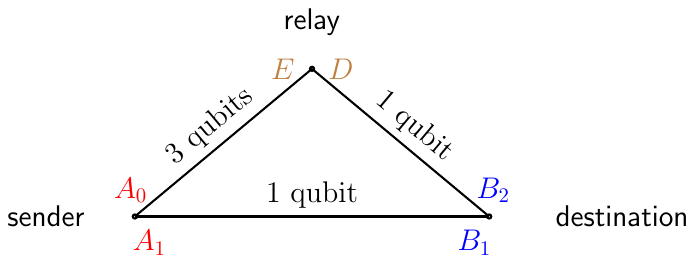} %
\end{center}
\caption{
A graphical representation for the noiseless relay channel in Example~\ref{Example:Trivial_AF}.
Alice can transmit 
three qubits to the relay, from $A_0$ to $E$,
and
a single qubit to Bob, from $A_1$ to $B_1$. The relay can send a single qubit to Bob, from $D$ to $B_2$.
}
\label{Figure:Id_Relay}
\end{figure}

We give two examples to illustrate our results. 
\begin{example}[Wired network]
\label{Example:Trivial_AF}
We begin with  a trivial example, where the assist-forward strategy achieves capacity.
The model below is appropriate for a \emph{wired network} of optical fibers \cite{Kramer:08n}. 
Suppose that Alice and Bob's systems $A$ and $B$ each comprises two components,
$A=(A_0,A_1)$ and $B=(B_1,B_2)$. 
We consider a noiseless relay channel with a graphical representation as in Figure~\ref{Figure:Id_Relay},   
\begin{align}
\mathcal{N}_{A_0 A_1 D\to B_1 B_2 E}= \mathrm{id}_{A_1\to B_1}\otimes \mathrm{id}_{D\to B_2}\otimes \mathrm{id}_{A_0\to E}
\end{align}
 where
$A_1$, $B_1$, $D$ and $B_2$ are qubits,  and
$A_0$ and $E$ are of dimension eight (each consists of three qubits). 
This yields the capacity value
\begin{align}
C(\mathcal{N})=
\mathsf{R}_{\text{A-F}}(\mathcal{N})=2 \,.
\end{align}
Using the assist-forward approach, Alice sends her $2$-bit message to the relay and also generates an EPR pair between the relay and Bob. Subsequently, the relay communicates the message to Bob using the superdense coding protocol.

Of course, the assist-forward strategy is completely unnecessary in this case. Instead,
we can use a partial decode-forward scheme and
 simply send one bit through the relay and one bit through the direct channel to Bob. 

\end{example}

\begin{example}[Depolarizing relay channel]
\label{Example:Depolarizing}
Consider a quantum relay channel with ORC, $B_1$ and $B_2$, such that 
\begin{align}
\mathcal{N}_{AD\to B_1 B_2 E}&=
\mathcal{M}_{A\to B_1 E}\otimes \mathcal{P}_{D\to B_2} 
\,,
\intertext{where $A$, $D$, $B_1$, $B_2$, $E$ are qubits, and}
\mathcal{M}_{A\to B_1 E}(\rho)&=
\frac{1}{4}\Big[
\rho\otimes \theta_0+
(\mathsf{X}\otimes \mathsf{X})(\rho\otimes \theta_0)(\mathsf{X}\otimes \mathsf{X})
\nonumber\\&
+
(\mathsf{Y}\otimes \mathsf{Y})(\rho\otimes \theta_0)(\mathsf{Y}\otimes \mathsf{Y})
+
(\mathsf{Z}\otimes \mathsf{Z})(\rho\otimes \theta_0)(\mathsf{Z}\otimes \mathsf{Z})
\Big]
\,,
\\
\mathcal{P}_{D\to B_2}(\rho)&=
(1-q)\rho+
q\frac{\identity}{2}
\,,
\label{Equation:Ortho_Receiver_Depolarizing}
\intertext{with}
\theta_0&=(1-p)\ketbra{0}+p\ketbra{1} 
\end{align}
where $\mathsf{X}$, $\mathsf{Y}$, $\mathsf{Z}$ denote the Pauli operators, and $p,q\in [0,1]$ are given parameters.
We note that the marginal channels
$\mathcal{M}_{A\to B_1}$ and $\mathcal{L}_{A\to E}$, to Bob and the relay, respectively, are both completely depolarizing channels, i.e., 
$\mathcal{M}_{A\to B_1}(\rho)=\mathcal{L}_{A\to E}(\rho)=\frac{\identity}{2}$ for every qubit state $\rho\in\mathfrak{D}(\mathcal{H}_A)$, by the Pauli twirl identity \cite[Ex. 4.7.3]{Wilde:17b}. Based on the measure-forward bound in Theorem~\ref{theo:MF}, we show that the capacity of the depolarizing relay channel satisfies
\begin{align}
C(\mathcal{N})\geq 1-h\left(p*\frac{q}{2}\right)
\end{align}
where $h(t)=-(1-t)\log(1-t)-t\log(t)$ is the binary entropy function, and 
$\alpha*\beta=(1-\alpha)\beta+\alpha(1-\beta)$.
The analysis is given in Appendix~\ref{Appendix:Depolarizing}.
Clearly, without the relays help, the capacity would have been zero.
\end{example}

\section{Summary and Discussion}
\label{Section:Summary}

\subsection{Quantum relay channel}
We  consider the fully quantum relay channel $\mathcal{N}_{AD\to BE}$, where Alice transmits $A$, the relay transmits $D$, Bob receives $B$, and the relay receives $E$. 
When there are multiple rounds of communication,  synchronization is important. 
Suppose that the operation is governed by a central clock that ticks $n$ times.
Between the clock ticks $i-1$ and $i$, the sender and the relay transmit the channel inputs $A_i$ and $D_i$, respectively. See Figure~\ref{Figure:Q_Relay_Coding}. 
Then, at clock tick $i$,
the relay and destination receivers receive the channel outputs $E_i$ and $B_i$, respectively. This requires a small delay before reception to ensure causality.
The relay is only required to assist the transmission of information to Bob, and does not necessarily decode any information (see Remark~\ref{Remark:Relay_Decoding}). 
The classical channel model was originally introduced by van der Meulen \cite{vanderMeulen:71p} as a building block for multihop networks \cite{PeregSteinberg:19p3}.
Here, we consider the quantum counterpart.

Our coding definitions in Section~\ref{Section:Coding} are  more involved than in the classical setting. %
Since the channel is fully quantum, we need to account for the post-operation state at the relay in each time instance (see Remark~\ref{Remark:Leftover}). That is, the relay encoding at time $i$ affects the state of $E_{i-1}$, as well as the previous outputs. 
The system $\bar{E}^{i-1}$ in Section~\ref{Section:Coding} can thus be viewed as a ``leftover" of the relay encoding at time $i$. This leftover can be used in subsequent steps, at time 
$j>i$. For example, if the relay performs a sequence of measurements, then at time $i$, the relay applies a quantum instrument such that $\bar{E}^{i-1}$ is the post-measurement system, while $D_i$ stores the measurement outcome.
In the classical setting, $\bar{E}^{i-1}$ is simply a copy of the previously received observations at the relay.

\subsection{Achievability results}
We  establish three bounds that are based on different coding strategies, i.e., partial decode-forward, measure-forward, and assist-forward.
Using the partial decode-forward strategy,
the relay decodes part of the information, while the other part is decoded without the relay's help (Section~\ref{Subsection:PDF}).
Based on the partial 
decode-forward bound,
we determine the capacity for the special class of Hadamard relay channels. We also recover the result by Savov et al. \cite{SavovWildeVu:12c} for  the special case of a classical-quantum relay channel. Furthermore, we observed that when the relay receives Bob's entire environment, our results yield a bound that resembles the environment-assisted distillation rate \cite{Smolin:05p,Winter:05a}.

The formula for the partial decode-forward bound includes three random variables,
$X_0$, $X_1$, and $U$. Intuitively, the auxiliary variables are associated with the information sent by the sender, the information sent by the relay, and the information that is decoded by the relay, respectively (see Remark~\ref{Remark:External_U}).
Taking $U$ to be null, we obtain the direct-transmission lower bound, i.e., 
 the Holevo information (see Remark~\ref{Remark:Direct_Full_DF}).
In the case of an anti-degraded classical-quantum channel, direct transmission is optimal. 
This capacity result is intuitive, since Bob's channel is better than the relay's in this trivial case (see Remark~\ref{Remark:Anti_Degradable}). 
On the other hand, taking $U=X_0$ implies that the relay decodes the entire message, which corresponds to a full decode-forward strategy (see \eqref{eq:low_Full_DF}).
This turns out to be optimal for Hadamard relay channels, as established in Thoerem~\ref{Theorem:Hadamard_Relay}.

In the  measure-forward coding scheme, the relay does not decode information at all (Section~\ref{Subsection:MF}). Instead, the relay performs a sequence of measurements, and then sends a compressed representation of the measurement outcome. This generalizes the classical compress-forward bound \cite{CoverElGamal:79p}. 
We interpret
the formula %
as follows. 
The encoder and the relay prepare their input states using input ensembles that are indexed by 
$X_0$ and $X_1$, respectively.
The relay receives $E$ and performs a measurement, which yields $Y_1$ as an outcome. 
The measurement outcome is then compressed and encoded by $Z_1$. Bob receives the output $B$.
Roughly speaking,     the number of information bits that are obtained from the measurement compression operation is $n I(Z_1;Y_1|X_1 B)_\omega$, while  the number of information bits that the relay can send through the channel to Bob is below $nI(X_1;B)_\rho$. This limitation is reflected through the maximization constraint, $I(Z_1;Y_1|X_1 B)_\omega\leq I(X_1;B)_\rho$ (see Remark~\ref{Remark:MF}).

At last, we consider quantum relay channels with orthogonal receiver components (Section~\ref{Subsection:AF}).
In this case, Bob's output has two components
$B_1$ and $B_2$, and
the quantum relay channel is a product of two channels, 
 a broadcast channel from the sender to the relay and the receiver component $B_1$, and a direct channel  from the relay to the receiver component $B_2$. 
 The relay channel is referred to as having orthogonal receiver components due to this decoupling (see Remark~\ref{Remark:ORC}). 
The assist-forward bound is based on a new approach, whereby the transmitter sends the message to the relay and simultaneously generates entanglement assistance between the relay and the destination receiver. Subsequently, the relay can transmit the message to the destination receiver with rate-limited entanglement assistance.
We interpret 
the formula for the assist-forward bound  as follows. 
The input to the broadcast channel is chosen from an ensemble  $\{\theta^{x_1}\}$. 
Alice can send the message to the relay in this manner at rate %
$R<I(X_1;E)_\theta$.
Alice also generates entangled pairs, $G_0^n$ and $G_1^n$.
She simultaneously uses the broadcast channel to distribute the entanglement between Bob and the relay, respectively.
The entanglement rate is roughly $Q= \min \left\{
I(G_0 \rangle B_1 X_1 )_\theta
\,,\;
I(G_1 \rangle E X_1 )_\theta 
\right\}$.
In the subsequent block, the relay and Bob perform an entanglement-assisted communication protocol, using the entanglement resources that were generated in the previous block. 
This requires %
both 
$R<I(X_2 G_2;B_2)_\zeta$ and $R< I(X_2;B_2)_\zeta+ I(G_2\rangle B_2 X_2)_\zeta +Q$, based on Shor's result \cite{Shor:04p} on rate-limited entanglement assistance (see Remark~\ref{Remark:AF}).

Section~\ref{Section:Examples} demonstrates our results through two examples.
The first is a trivial example that represents a wired network, where the links are noiseless. 
We observe that both the partial decode-forward and assist-forward are capacity achieving in this case. 
Furthermore, we use our measure-forward result in order to compute an achievable rate for a depolarizing relay channel with orthogonal receiver components.

A shortcoming of our results, as well as the previous results on the c-q relay channel  \cite{SavovWildeVu:12c,Savov:12z,DingGharibyanHaydenWalter:20p} ,
is that we do not have a bound on the alphabet cardinality for the auxiliary random variables and the dimension of the auxiliary systems: $U$, $X_0$, $X_1$ in the partial decode-forward bound,  
$X_0$, $X_1$, $Y_1$, $Z_1$ in the measure-forward bound, and
$X_1$, $X_2$, $G_0$, $G_1$ in the assist-forward bound. 
Although one can always compute an achievable rate by simply choosing the dimensions, the best lower bound cannot be 
 computed exactly in general.
A similar difficulty appears in other quantum models such as the  broadcast channel (see Discussion section in \cite{DupuisHaydenLi:10p}), wiretap channel \cite[Remark 5]{QiSharmaWilde:18p}, masking
\cite[Remark 10]{PeregDeppeBoche:21p3}, 
and squashed entanglement \cite[Section 1]{LiWinter:14p}.

\subsection{Degraded channels}
\label{Discussion:Degraded}
In many networks settings, both in classical and quantum  information theory, degraded channels are understood a lot better than general channels.
We have established a single-letter formula for the class of Hadamard relay channels, i.e., a  measure-and-prepare channel that is %
assumed to be degraded. 

Recall that a degraded channel is defined as follows. 
We say that a quantum relay channel $\mathcal{N}_{A\to BE}$ is degraded if there exists a degrading channel
$\mathcal{P}_{E\to BE}$ such that 
\begin{align}
\mathcal{N}_{A\to BE}=\mathcal{P}_{E\to BE}\circ
\mathcal{M}_{A\to E} \,.
\label{Equation:Discussion_Physically_Degraded}
\end{align}
(see Definition~\ref{Definition:Degraded}). Intuitively, if a relay channel is degraded, then  the output state of the destination receiver is a noisy version of that of the relay.

We may also write the definition in terms of the Choi state. In general, a quantum channel $\mathcal{T}_{A\to B}$ is fully characterized by the Choi state
$%
\xi_{A_1 B}\equiv (\mathrm{id}_{A_1}\otimes \mathcal{T}_{A\to B}) (\ketbra{\Phi}_{A_1 A})
$, %
where $\ket{\Phi}_{A_1 A}=\frac{1}{\sqrt{\mathrm{dim}(\mathcal{H}_A)}}\sum_{j=0}^{\mathrm{dim}(\mathcal{H}_A)-1} \ket{j}\otimes \ket{j}$ is a maximally entangled state. The quantum relay channel
$\mathcal{N}_{A\to BE}$ is thus characterized by a tripartite Choi state, $\xi_{A_1 BE}$.
Therefore, the quantum relay channel $\mathcal{N}_{A\to BE}$ is degraded if and only if its Choi state forms a quantum Markov chain 
\begin{align}
A_1\Cbar E\Cbar B
\end{align}
(see  \cite[Def. 5.1]{Sutter:18b}).
That is, there exists a recovery channel $\mathcal{P}_{E\to BE}$ such that  
\begin{align}
\xi_{A_1 BE}=(\mathrm{id}_{A_1}
\otimes \mathcal{P}_{E\to BE})
(\xi_{A_1 E}) \,.
\end{align}
This holds if and only if $I(A_1;B|E)_\xi=0$ (see  \cite[Th. 5.2]{Sutter:18b}).

There are various notions of degradedness in the literature \cite{Steinberg:25a}.
In the classical literature, a classical channel that satisfies the property in \eqref{Equation:Discussion_Physically_Degraded} above %
is called a 
\emph{physically} degraded channel.
In broadcast and security settings, the capacity characterization only depends on the marginals.
Therefore, it suffices to require that there exists a
degrading channel $\widetilde{\mathcal{P}}_{E\rightarrow B}$ such that the marginals satisfy the following relation:
	\begin{align}
	\mathcal{L}_{AD\to B}=\widetilde{\mathcal{P}}_{E\to B}\circ\mathcal{M}_{AD\to E} \,.
    \label{Equation:Discussion_Stochastically_Degraded}
	\end{align}
A channel that satisfies \eqref{Equation:Discussion_Stochastically_Degraded} is referred to as a \emph{stochastically} degraded channel.
More general classes of degraded channels include the ``less noisy" and ``more capable" channels. 
In particular, a broadcast channel $\mathcal{N}_{A\to BE}$ is called less noisy if $I(X;B^n)_\rho\leq I(X;E^n)_\rho$ for every pure state ensemble
$\{ p_X, \rho_{A^n}^x \}$ for $\mathcal{N}_{A\to BE}^{\otimes n}$, $n=1,2,\ldots$
(see \cite[Def. 2]{Watanabe:12p}). 
Steinberg~\cite{Steinberg:25a} has also introduced the classical class of  $\eta$-less noisy channel, for which
$I(X;B^n)_\rho\leq \eta I(X;E^n)_\rho$, for some $\eta\in (0,1)$.

In the classical setting, the capacity of the physically degraded relay channel is known, as the decode-forward lower bound and the cutset upper bound coincide. Recently, Steinberg \cite{Steinberg:25a} established a full capacity characterization for a  classical family of $\eta$-less noisy and 
\emph{primitive} relay channels, where there is a noiseless bit pipe from the relay to the destination receiver \cite{PeregDeppeBoche:21p2}.  
The primitive relay channel can be viewed  
as a simplified version of the relay channel with orthogonal receiver components (see Subsection~\ref{Subsection:AF}).

 \subsection{Quantum Repeaters}
Attenuation in optical fibers poses a significant challenge for long-distance quantum communication protocols. This limitation affects current applications such as quantum key distribution \cite{JTNMVL:09p}, and future development of  the quantum Internet \cite{BSDP:19c} and quantum networks more broadly \cite{BFSDBFJ:20b}. Quantum repeaters have emerged as a promising solution by acting as a relay of quantum information  \cite{BriegelDurCiracZoller:98p}.
Here, we addressed the classical task of sending messages through the channel.  
In order to distribute entanglement and send quantum information, the quantum relay would need to operate as a quantum repeater \cite{BriegelDurCiracZoller:98p}.  
 In the model's simplest form, the sender employs quantum communication to the relay (repeater) in order to generate an EPR  pair $\ket{\Phi_{A E}}$ between Alice and the relay. At the same time, the relay generates an EPR pair $\ket{\Phi_{D B}}$ with the destination receiver.
 Then, the repeater can perform a Bell measurement on $E,D$, thus swapping the entanglement such that $A$ and $B$ are now entangled.
Pereg et al. \cite{PeregDeppeBoche:21p2} provided an information-theoretic perspective on quantum repeaters through the task of
 quantum subspace transmission via a primitive relay channel, with a noiseless qubit pipe from the relay to the destination receiver.
The general case remains unsolved.

\section*{Acknowledgments}
The author wishes to thank Gerhard Kramer (Technical University of Munich) for useful discussions.
 The author was supported in part by
Israel Science Foundation (ISF) under grants 939/23 and 2691/23, in part by
German–Israeli Project Cooperation (DIP) within the Deutsche Forschungsgemeinschaft (DFG) under grant 2032991, in part by the Ollendorff Minerva
Center (OMC) of Technion under grant 86160946, in part by the Nevet
Program of the Helen Diller Quantum Center at Technion under grant
2033613.

\begin{appendices} %
{

\section{Information Theoretic Tools}
\label{App:Tools}
Our analysis  builds on the quantum  method of types. The basic definitions and lemmas that are used in the analysis are given below.

\subsection{Typical Projectors}
\label{App:Q_Typical}
We begin with the classical typical set.
The type of a classical sequence $x^n\in\mathcal{X}^n$ is defined as the empirical distribution $\hat{P}_{x^n}(a)=N(a|x^n)/n$ for $a\in\mathcal{X}$, where $N(a|x^n)$ is the number of occurrences of the letter $a\in\mathcal{X}$ in the sequence $x^n$.
Let $\delta>0$.
The $\delta$-typical set with respect to a probability distribution $p_X$ is defined as the following set of sequences,
\begin{align}
T_\delta^{(n)}(p_X)=
\left\{
x^n\in\mathcal{X}^n \,:\;
\abs{
p_X(a)-\hat{P}_{x^n}(a)
}\leq \delta p_X(a)
\,,\;
\text{for all $a\in\mathcal{X}$}
\right\}
\,.
\end{align}
If $p_X$ is a type on $\mathcal{X}^n$, then $\mathcal{T}^{(n)}(p_X)$ denotes the corresponding type class, i.e., 
\begin{align}
\mathcal{T}^{(n)}(p_X)=
\left\{
x^n\in\mathcal{X}^n \,:\;
\hat{P}_{x^n}(a)= p_X(a)
\,,\;
\text{for all $a\in\mathcal{X}$}
\right\}
\,.
\end{align}
For a pair of sequences $x^n$ and $y^n$, we give similar definitions in terms of the joint type $\hat{P}_{x^n,y^n}(a,b)=N(a,b|x^n,y^n)/n$ for $a\in\mathcal{X}$, $b\in\mathcal{Y}$, where $N(a,b|x^n,y^n)$ is the number of occurrences of the symbol pair $(a,b)$ in the sequence 
$(x_i,y_i)_{i=1}^n$.

We move to the quantum method of types. 
Consider an ensemble $\{ p_X(x), \ket{x} \}_{x\in\mathcal{X}}$, with an average density operator,
\begin{align}
\rho=\sum_{x\in\mathcal{X}} p_X(x) \ketbra{x} \,.
\end{align}
The $\delta$-typical projector 
$\Pi_\delta^{(n)}(\rho)$ is an operator that projects onto the subspace spanned by vectors $\ket{x^n}$ that correspond to $\delta$-typical sequences. 
Specifically,
\begin{align}
\Pi_\delta^{(n)}(\rho)\equiv \sum_{x^n\in
T_\delta^{(n)}(p_X)} \ketbra{ x^n } 
\label{eq:ProjType}
\end{align}
with $\ket{x^n}= \bigotimes_{i=1}^n\ket{x_i}$.
Let
$0<\delta<\nu_X$, where 
$\nu_X\equiv 
\min\{p_X(a)\,:\;
p_X(a)>0\}$.
Based on the classical typicality properties \cite[Th. 1.1]{Kramer:08n}, 
\begin{align}
 \trace\left\{ 
 \Pi_\delta^{(n)}(\rho) 
\rho^{\otimes n} \right\}
&\geq 1-
\varepsilon_\delta(n)
\,,\;
\label{eq:UnitT} 
\\
\trace\left\{ \Pi_\delta^{(n)}(\rho) \right\}&\leq 2^{n(1+\delta)H(\rho)}
\,,\;
\\
\Pi_\delta^{(n)}(\rho)\cdot \rho^{\otimes n} \cdot \Pi_\delta^{(n)}(\rho) &\leq 2^{-n(1-\delta)H(\rho)}\cdot  \Pi_\delta^{(n)}(\rho)
\end{align}
where $\varepsilon_\delta(n)=2d_A e^{-n\nu_X\delta^2}$.

We will also need conditional typicality.
Consider an ensemble $\{ p_X(x), \rho_x \}_{x\in\mathcal{X}}$ in
$\mathfrak{D}(\mathcal{H}_B)$, with an average classical-quantum state
\begin{align}
\omega_{XB}=\sum_{a\in\mathcal{X}} 
p_X(a) \ketbra{a}_X\otimes  \rho_a \,.
\end{align}
Each state in the ensemble has a spectral decomposition
$\rho_a=\sum_{b\in\mathcal{Y}} p_{Y|X}(b|a)\ketbra{\psi_{a,b}}$, for $a\in\mathcal{X}$.
Given a fixed sequence $x^n\in\mathcal{X}^n$, divide the index set $[1:n]$ into the subsets $I_n(a)=\{ i: x_i=a  \}$, $a\in\mathcal{X}$.
The conditional $\delta$-typical projector $\Pi_\delta^{(n)}(\omega|x^n)$ is defined such that for every $a\in\mathcal{X}$, we apply 
the $\delta$-typical projector with respect to $\rho_a$ on the systems 
$\{ B_i \,:\; x_i=a \} $. Specifically,
\begin{align}
\Pi_\delta^{(n)}(\omega_{XB}|x^n)\equiv \bigotimes_{a\in\mathcal{X}}  %
\Pi_\delta^{(N(a|x^n))}(\rho_a)
\,.
\end{align}
Fix $0<\delta_1<\delta<\nu_{XY}$ and 
$x^n\in T_{\delta_1}(p_X)$.
Based on the classical typicality properties \cite[Th. 1.2]{Kramer:08n}, 
\begin{align}
 \trace\left\{ 
 \Pi_\delta^{(n)}(\omega_{XB}|x^n) 
\rho_{x^ n} \right\}
&\geq 1-
\varepsilon_{\delta_1,\delta}(n)
\,,\;
\label{eq:UnitT_c_1} 
 \\
\trace\left\{ \Pi_\delta^{(n)}(\omega_{XB}|x^n) \right\}&\leq 2^{n(1+\delta)H(B|X)_\omega}
\,,\;
\\
\Pi_\delta^{(n)}(\omega_{XB}|x^n)\cdot \rho_{x^n} \cdot \Pi_\delta^{(n)}(\omega_{XB}|x^n) &\leq 2^{-n(1-\delta_1)H(B|X)_\omega}\cdot  \Pi_\delta^{(n)}(\omega|x^n)
\end{align}
where $\varepsilon_{\delta_1,\delta}(n)=1-\left[1- \varepsilon_{\delta}\left(n(1-\delta_1)\nu_X)\right)\right]^{d_A}$ and 
$\rho_{x^ n}\equiv \bigotimes_{i=1}^n \rho_{x_i}$. 
Furthermore, 
\begin{align}
\trace( \Pi^\delta(\omega_B) \rho_{x^n} )\geq& 1-\varepsilon_{\delta_1,\delta}(n)
\label{eq:UnitTCondB}
\end{align}
 (see \cite[Property 15.2.7]{Wilde:17b}). 

In a random coding scheme,
it is often useful to limit the codewords to the $\delta$-typical set. Given a probability distribution 
$p_{X}$ on $\mathcal{X}$, denote
the $n$-fold product by 
$p^n_{X}$, i.e.,
\begin{align}
p^n_{X}(x^n)\equiv \prod_{i=1}^n p_{X}(x_i) \,,\, \text{ for 
$x^n\in\mathcal{X}^n$.}
\end{align}
Then, 
define a distribution $\bar{p}_{X^n}$ by %
\begin{align}
\bar{p}_{X^n}(x^n)=
\begin{cases}
p_X^n(x^n)/
\sum\limits_{x'^n\in T_\delta(p_X)}p_X^n(x'^n)
&\text{if $x^n\in T_\delta(p_X)$}
\\
0 &\text{if $x^n\notin T_\delta(p_X)$}
\end{cases}
\end{align}
and let 
\begin{align}
\bar{\rho}_{B^n}=
\sum_{x^n\in T_\delta^{(n)}(p_X)} \bar{p}_{X^n}(x^n)\rho_{x^n}
\,.
\end{align}
Then, we have the following properties
\begin{align}
\sum_{x^n\in\mathcal{X}^n}
\abs{\bar{p}_{X^n}(x^n)-p_X^n(x^n)}&\leq 2\varepsilon_\delta(n)
\,,
\\
\Pi_\delta^{(n)}(\omega_B)
\cdot \bar{\rho}_{B^n} \cdot
\Pi_\delta^{(n)}(\omega_B)
&\leq 
2^{-n\left( (1-\delta)H(B)_\omega-\bar{\varepsilon}_\delta(n) \right)}
\Pi_\delta^{(n)}(\omega_B)
\,.
\label{eq:PidimCond_bar}
\end{align}
where $\bar{\varepsilon}_\delta(n)=-\frac{1}{n}\log[1-\varepsilon_\delta(n)]$
(see \cite[Ex. 20.3.2]{Wilde:17b}).
Notice that  
 $\varepsilon_\delta(n)$, $\bar{\varepsilon}_\delta(n)$, 
 $\varepsilon_{\delta_1,\delta}(n)$ all tend to zero as $n\to\infty$. 
A joint distribution 
$\bar{p}_{X^n Y^n}$ is defined in the same manner, with respect to the joint typical set.

 \subsection{Quantum Packing and Gentle Measurement Lemmas}
 The quantum packing lemma is a useful tool in achievability proofs. Consider the following one-shot communication setting.
Suppose that Alice has a classical codebook that consists  of $\mathsf{M}$ codewords. Given a message $m\in [1:\mathsf{M}]$,  she sends a codeword $x(m)$ through a classical-quantum channel, producing an output state $\rho_{x(m)}$ at the receiver. The  quantum packing lemma provides a decoding measurement in order for Bob to recover $m$, by using a random codebook.
The proof is based on the square-root measurement \cite{Holevo:98p,SchumacherWestmoreland:97p}.

\begin{lemma}[Quantum Packing Lemma  \cite{HsiehDevetakWinter:08p}]
\label{lemm:Qpacking}
Consider an ensemble,
$\{ p_X(x), \rho_{x}\}_{x\in\mathcal{X}}$, with an average state,
\begin{align}
\rho=\sum_{x\in\mathcal{X}} p_X(x) \rho_{x}
 \,.
\end{align} 
Furthermore, suppose that there exist  a code projector $\Pi$ and codeword projectors $\Pi_{x}$, $x\in \mathcal{X}$, that satisfy the following conditions:
\begin{align}
\trace(\Pi \rho_{x})&\geq\, 1-\varepsilon 
\label{Equation:Condition_1}\\
\trace(\Pi_{x}\rho_{x})&\geq\, 1-\varepsilon 
\label{Equation:Condition_2}\\
\trace(\Pi_{x})&\leq\, 2^{\mathsf{h}}
\label{Equation:Condition_3}\\
\Pi \rho \Pi &\leq\, 2^{-\mathsf{H}} \Omega
\label{Equation:Condition_4}
\end{align}
for all $x\in\mathcal{X}$ and some $\varepsilon \in (0,1)$, $0<\mathsf{h}<\mathsf{H}$.
Let $\mathscr{A}=\{X(m)\}_{m\in [1:\mathsf{M}]}$, be a random codebook of size $\mathsf{M}$, where the codewords are drawn independently at random, according to $p_X$.
Then, there exists a decoding POVM
 $\{ \Lambda_m \}_{m\in [1:\mathsf{M}]}$ such that the expected probability of error satisfies
\begin{align}
\label{eq:QpackB}
  \mathbb{E}_{\mathscr{A}}\left[
  1-\frac{1}{\mathsf{M}}\sum_{m=1}^{\mathsf{M}}\trace\left\{ \Lambda_m\cdot  \rho_{X(m)} \right\} \right]  \leq 
2(\varepsilon +2\sqrt{\varepsilon })
+
4\mathsf{M}\cdot
2^{-(\mathsf{H}-\mathsf{h})}
\end{align} 
where the expectation is with respect to the random codebook, $\mathscr{A}=\{X(m)\}$.
\end{lemma}

The gentle measurement lemma is  useful in our analysis, since it guarantees that we can perform multiple measurements such that the 
state of the system remains almost the same after each measurement (see also \cite{Pereg:21p}).

\begin{lemma}[Gentle Measurement Lemma  {\cite{Winter:99p,OgawaNagaoka:07p}}]
\label{lemm:gentleM}
Let $\rho$ be a density operator. Suppose that $\Lambda$ is a measurement operator such that $0\leq \Lambda\leq \identity$. If
\begin{align}
\trace(\Lambda\rho) \geq 1-\delta
\end{align}
for some $0\leq\delta\leq 1$, then the post-measurement state $\tilde{\rho}\equiv \frac{\sqrt{\Lambda}\rho\sqrt{\Lambda} }{\trace(\Lambda\rho)}$ is $2\sqrt{\delta}$-close to the original state in trace distance, i.e.
\begin{align}
\norm{ \rho-\tilde{\rho} }_1\leq 2\sqrt{\delta} \,.
\end{align}
\end{lemma}
In our analysis, we will establish the conditions of the gentle measurement lemma 
based on the property \eqref{eq:QpackB}, arising from the quantum packing lemma.

\section{Proof of Theorem~\ref{theo:PDF} (Partial Decode-Forward Strategy)}
\label{Appendix:PDF}
Consider a   quantum relay channel $\mathcal{N}_{AD\to BE}$. 
The proof extends the classical scheme of partial decode-forward coding, along with observations from a previous work by the  author \cite{Pereg:21p}.

We show that for every $\varepsilon_0,\delta_0>0$, there exists a $(2^{n(R-\delta_0)},n,\varepsilon_0)$ code for the quantum relay channel $\mathcal{N}_{DA\to BE}$, provided that $R< \mathsf{R}_{\text{PD-F}}(\mathcal{N})$. 
To prove achievability, we extend  the classical block Markov coding to the quantum setting, and then apply the quantum packing lemma and the classical covering lemma. We use the gentle measurement lemma \cite{Winter:99p}, %
which guarantees that multiple decoding measurements can be performed without ``destroying" the output state.

Fix a given input ensemble, $\{ p_{U}(u)p_{X_0 X_1|U}(x_0,x_1|u) , \theta_{A}^{x_0}\otimes
\zeta_D^{x_1}\}$. Denote the output states by 
\begin{align}
\omega_{BE}^{x_0,x_1}&\equiv \mathcal{N}_{AD\to B E}(\theta_{A}^{x_0}\otimes
\zeta_D^{x_1}) \,,
\end{align}
and
the average states,
\begin{align}
\omega_{AD}^{u,x_0}&\equiv
\theta_A^{x_0}\otimes \sum_{x_1\in\mathcal{X}_1} p_{X_1|X_0 U}(x_1|x_0,u) \zeta_D^{x_1} \,,\;
\omega_{BE}^{u,x_0}\equiv
\mathcal{N}_{AD\to B E}(\omega_{AD}^{u,x_0}) \,,
\\
\omega_{AD}^{u,x_1}&\equiv
\sum_{x_0\in\mathcal{X}_0} p_{X_0|X_1 U}(x_0|x_1,u) \theta_A^{x_0}\otimes \zeta_D^{x_1} \,,\;
\omega_{BE}^{u,x_1}\equiv
\mathcal{N}_{AD\to B E}(\omega_{AD}^{u,x_1}) \,,
\\
\omega_{BE}&\equiv
\sum_{u\in\mathcal{U}} p_U(u)
\sum_{x_0\in\mathcal{X}_0}
\sum_{x_1\in\mathcal{X}_1} p_{X_0 X_1|U}(x_0,x_1|u) \omega_{BE}^{x_0,x_1} \,,
\end{align}
for $(u,x_0,x_1)\in\mathcal{U}\times\mathcal{X}_0
\times \mathcal{X}_1$.

Recall that the relay encodes in a strictly-causal manner. Specifically, the relay  has access to the sequence of previously received systems, $E_{i-1},\bar{E}^{i-2}$, from the past.
We use $T$ transmission blocks, where each block consists of $n$ input systems. In particular,  the relay has access to the systems from the previous blocks.
In effect, the $j^{\text{th}}$ transmission block of the relay encodes part of the message $m_{j-1}\in [1:2^{nR}]$ from the previous block. 

First,
we use rate splitting.
Let every message $m_j$, $j\in [1:T-1]$,
comprise two independent components $m_j'$ and $m_j''$, where 
$m_j'\in [1:2^{nR'}]$ and
$m_j''\in [1:2^{nR''}]$, such that 
$R=R'+R''$.
The coding scheme is referred to as a ``partial decode-forward" strategy, since 
the relay  decodes the first component alone.

\subsection{Code Construction}

The code construction, encoding and decoding procedures are described below.
We illustrate the code structure in Figure~\ref{fig:PDF}.

\begin{figure*}

\begin{tabular}{l|ccccc}
{\small Block}				
& $1$ & $2$ & $\cdots$ & $T-1$ & $T$ \\
								
\\ \hline &&&&& \\ 
$U$	& $u^n(m_1'|1)$ &	$u^n(m_2'|m_1')$	& $\cdots$ & $u^n(m_{T-1}'|m_{T-2}'))$ & $u^n(1|m_{T-1}'))$ 
\\&&&&& \\
$A$ & $x_0^n(m_1',m_1''|1)$	&	$x_0^n(m_2',m_2''|m_1')$ & $\cdots$	& $x_0^n(m_{T-1}',m_{T-1}''|m_{T-2}')$ &  $x_0^n(1,m_1''|m_{T-1}')$ 
\\&&&&& \\
$E$	& $\tilde{m}_1' \rightarrow$ & $\tilde{m}_2' \rightarrow$	& $\cdots$	& $\tilde{m}_{T-1}' \rightarrow$ & $\emptyset$
\\&&&&& \\
$D$	& $x_1^n(1)$ & $x_1^n(\tilde{m}_1')$ & $\cdots$	& $x_1^n(\tilde{m}_{T-2}')$ &  $x_1^n(\tilde{m}_{T-1}')$
\\&&&&& \\
$B$ & $\emptyset$ & $\leftarrow\hat{m}_1'$ & $\cdots$ & $\leftarrow\hat{m}_{T-2}'$ & $\leftarrow\hat{m}_{T-1}'$
\\
& $\hat{m}_1''$ & $\hat{m}_2''$ & $\cdots$ & $\hat{m}_{T-1}''$ & $\emptyset$
\end{tabular}
\caption[Partial decode-forward strategy]{Partial decode-forward strategy. The block index $j\in [1:T]$ is indicated at the top. In the following rows, we have the corresponding elements: 
(1) auxiliary sequences; 
(2) codewords of Alice;  
(3) relay estimates;
(4) relay codewords;
(5), (6) estimated messages at the destination receiver.
The arrows in the third row indicate that the relay measures and encodes forward with respect to the block index, while the arrows in the fifth row indicate that Bob decodes backwards.  
}
\label{fig:PDF}
\end{figure*}

\subsubsection{Classical Codebook Construction}
For every $j\in [1:T]$, generate a classical codebook
$\mathscr{B}(j)$ as follows.
Select $(2^{n R'})^2$ independent sequences
$u^n(m_j'|m_{j-1}')$, for 
$m_j',m_{j-1}'\in [1:2^{nR'}]$, according to 
$\bar{p}_{U^n}$ (see the first row in Figure~\ref{fig:PDF}).
Then, for every given 
$u^n(m_j'|m_{j-1}')$,
select conditionally independent 
$(x_0^n(m_j',m_j''|m_{j-1}'),x_1^n(m_{j-1}'))$,
for 
$m_j''\in [1:2^{nR''}]$, according to 
$ \bar{p}_{X_0^n X_1^n|U^n}(\cdot,\cdot|u^n)$, conditioned on $u^n\equiv u^n(m_j'|m_{j-1}')$.

\subsubsection{Encoding}
Set $m_0'=m_T'=m_T''\equiv 1$.
Given the message sequence 
$(m_j',m_j'')_{j\in [1:T]}$,
prepare the input state
$\bigotimes_{j=1}^T \rho_{A^n(j)}$, where
\begin{align}
\rho_{A^n(j)}
=
\bigotimes_{i=1}^n
\theta_A^{x_{0,i}}
\,\text{ for $x_{0}^n\equiv 
x_0^n(m_j',m_j''|m_{j-1}')$}
\,.
\end{align}
Then, transmit $A^n(j)$ in Block $j$, for $j=1,2,\ldots,T$.
Hence, we encode in an average rate of $\left( \frac{T-1}{T} \right)R$, which tends to $R$ as $T\to \infty$.

The encoding operation is illustrated in the second~row of Figure~\ref{fig:PDF}.

\subsubsection{Relay Encoding}
Set $\tilde{m}_0'\equiv 1$.
\begin{enumerate}[(i)]
\item
At the end of Block $j$,
find an estimate $\tilde{m}_j'$ by performing a measurement $\{\Gamma_{m_j'|x_1^n}\}_{m_j'\in [1:2^{nR'}]}$ for 
$x_1^n\equiv x_1^n(\tilde{m}_{j-1}')$, which will be specified later, on the received systems
$E^n(j)$, for 
$j=1,2,\ldots,T-1$.

\item
In block $j+1$, prepare the state 
\begin{align}
\rho_{D^n(j+1)}=
\bigotimes_{i=1}^n
\zeta_{D}^{x_{1,i}}
\,,\;
\text{for $x_1^n\equiv x_1^n(\tilde{m}_j')$}
\end{align}
using the classical codebook
$\mathscr{B}(j+1)$. Then, transmit $D^n(j+1)$.
The relay's decoding and encoding operations are illustrated in the third and fourth~rows in Figure~\ref{fig:PDF}.
\end{enumerate}
This results in the output state $\bigotimes_{j=1}^T \rho_{B^n(j) E^n(j)}$, where
\begin{align}
&\rho_{B^n(j) E^n(j)}=
\bigotimes_{i=1}^n
\omega_{BE}^{x_{0,i},x_{1,i}}
\,,\;
\nonumber\\
&\text{ for 
$x_{0}^n\equiv 
x_0^n(m_j',m_j''|m_{j-1}')$ and
$x_1^n\equiv x_1^n(\tilde{m}_j')$.}
\end{align}

\subsubsection{Decoding}
Bob receives the $T$ output blocks, $B^n(1),\ldots,B^n(T)$, and decodes as follows. 
\begin{enumerate}[(i)]
\item
Decoding $(m_j')_{j\in [1:T]}$ is performed backwards.
Set $\hat{m}_0'=\hat{m}_T'\equiv 1$.
For $j=T-1,T-2,\ldots 1$,
find an estimate 
$\hat{m}_j'$ by performing a measurement
$
\{\Delta^1_{m_j'|\hat{m}_{j+1}'}\}$, which will be specified later, on the output systems $B^n(j+1)$. See the fifth~row in  Figure~\ref{fig:PDF}.
\item
Next, we decode $(m_j'')_{j\in [1:T]}$ as follows. 
For $j=1,2,\ldots T-1$,
find an estimate 
$\hat{m}_j''$ by performing a second measurement
$
\{\Delta^2_{m_j''|u^n,x_1^n}\}$ for 
$u^n\equiv u^n(\hat{m}_j'|\hat{m}_{j-1}'')$ and $x_1^n\equiv x_1^n(\hat{m}_j')$
,  which will also be specified later, on  $B^n(j)$. See the last row in  Figure~\ref{fig:PDF}.
\end{enumerate}

\subsection{Error Analysis}
By symmetry, we may assume without loss of generality that Alice sends the messages $(M_j',M_j'')=(1,1)$, for $j\in [1:T]$.
Consider the error events,
\begin{align}
\mathscr{E}_0(j)&= \{  \widetilde{M}_j'\neq 1 \}\,, \\
\mathscr{E}_1(j)&= \{  \widehat{M}_j'\neq 1  \}\,,\\
\mathscr{E}_2(j)&= \{  \widehat{M}_j''\neq 1   \}\,.
\end{align}
The event $\mathscr{E}_0(j)$ is associated with erroneous decoding by the relay, and 
$\mathscr{E}_1(j)$, $\mathscr{E}_2(j)$ by the destination receiver.
By the union of events bound, the expected probability of error is bounded by
\begin{align}
\mathbb{E} P_{e}^{(Tn)}(\mathscr{C}) &\leq
\sum_{j=1}^{T-1} \Pr({ \mathscr{E}_0(j) }
)
+\sum_{j=1}^{T-1} \Pr({ \mathscr{E}_1(j) } | { 
\mathscr{E}_0^c(j)\cap \mathscr{E}_1^c(j+1) } )\nonumber\\&
+\sum_{j=1}^{T}\Pr({ \mathscr{E}_2(j) } | { 
\mathscr{E}_0^c(j)\cap \mathscr{E}_1^c(j) \cap \mathscr{E}_1^c(j-1) })
\label{eq:PeBsc}
\end{align}
where the conditioning on $M_j'=M_{j-1}'=M_j''=1$ is omitted for convenience of notation. 

To bound the first sum, we use the quantum packing lemma, Lemma~\ref{lemm:Qpacking}. 
Observe that based on the quantum typicality properties in Section~\ref{App:Q_Typical} of Appendix~\ref{App:Tools}, %
we have for sufficiently large $n$,
\begin{align}
\trace\left[ \Pi_{\delta}^{(n)}(\omega_{X_1 E}|x_1^n) \omega_{E^n}^{u^n,x_1^n} \right] &\geq 1-\delta
\\
\trace\left[ \Pi_{\delta}^{(n)}(\omega_{UX_1 E}|u^n,x_1^n) \omega_{E^n}^{u^n,x_1^n} \right] &\geq 1-\delta
\\
\trace\left[ \Pi_{\delta}^{(n)}(\omega_{U X_1 E}|u^n,x_1^n)  \right] &\leq 2^{ n(1+2\delta) H(E|U X_1)_{\omega}}  
\\
\Pi_{\delta}^{(n)}(\omega_{X_1 E}|x_1^n)  \omega^{x_1^n}_{E^n}   \Pi_{\delta}^{(n)}(\omega_{X_1 E}|x_1^n) &\leq 2^{ -n(1-2\delta)H(E|X_1)_{\omega} } \Pi_{\delta}^{(n)}(\omega_{E})
\end{align}
for all $(u^n,x_1^n)\in T^{(n)}_{\delta_1}(p_U p_{X_1|U})$ (see \eqref{eq:UnitT_c_1}-\eqref{eq:UnitTCondB}). 
Since the codebooks are statistically independent of each other, we have by Lemma~\ref{lemm:Qpacking} that there exists a POVM $\Gamma_{m_{j}'|x_1^n}$ such that
$%
\Pr({ \mathscr{E}_0(j) })  \leq 2^{ -n( I(U;E|X_1)_\omega -R'-\varepsilon_1) } 
$, %
which tends to zero as $n\rightarrow\infty$, provided that 
\begin{align}
R'< I(U;E|X_1)_\omega-\varepsilon_1  
\label{eq:B2}
\end{align}
for $\varepsilon_1=\varepsilon_1(\delta)\equiv 4\delta \log(d_E)$.

Moving to the third sum in the RHS of (\ref{eq:PeBsc}), suppose that $\mathscr{E}_1^c(j)\cap \mathscr{E}_2^c(j+1)$ occurred. Namely, the relay measured the correct $M_j'$, and the decoder recovered $M_{j+1}'$. 
Then, by the same packing lemma argument as given above, 
 there exists a POVM $\Delta^1_{m_{j}'|m_{j+1}'}$ such that
$%
\Pr({ \mathscr{E}_2(j) }|\mathscr{E}_1^c(j)\cap \mathscr{E}_2^c(j+1))  \leq 2^{ -n( I(UX_1;B)_\omega -R'-\varepsilon_2 ) } 
$, %
which tends to zero as $n\rightarrow\infty$ for
\begin{align}
R'< I(U X_1;B)_\omega -3\varepsilon_2
\label{eq:B2_Rp_PDF}
\end{align}
where $\varepsilon_2=\varepsilon_2(\delta)\equiv 4\delta \log( d_B)$.

Denote the state of the output systems $B^n(j)$ in Block $j$, after this measurement, by $\tilde{\rho}_{B^n(j)}$.
Then, observe that due to the packing lemma inequality (\ref{eq:QpackB}),  the gentle measurement lemma 
{\cite{Winter:99p,OgawaNagaoka:07p}}, Lemma~\ref{lemm:gentleM},
 implies that the post-measurement state is close to the original state in the sense that
\begin{align}
\frac{1}{2}\norm{\tilde{\rho}_{B^n(j)}-\rho_{B^n(j)}}_1 \leq 2^{ -n\frac{1}{2}( I(UX_1;B)_\omega -R'-\varepsilon_2 ) } \leq 2^{-\varepsilon_2 n}
\end{align}
for sufficiently large $n$ and rates as in (\ref{eq:B2_Rp_PDF}).
Therefore, the distribution of measurement outcomes when $\tilde{\rho}_{B^n(j)}$ is measured is roughly the same as if the POVM $\Delta^1_{m_j'|m_{j+1}'}$ was never performed. To be precise, the difference between the probability of a measurement outcome $\hat{m}_j''$ when $\tilde{\rho}_{B^n(j)}$ is measured and the probability when $\rho_{B^n(j)}$ is measured is bounded by $ 2^{-\varepsilon_2 n}$ in absolute value \cite[Lemma 9.11]{Wilde:17b}.
Furthermore, the $\delta$-typicality properties:
\begin{align}
\trace\left[ \Pi_{\delta}^{(n)}(\omega_{UX_1 B}|u^n,x_1^n) \omega_{B^n}^{u^n,x_0^n,x_1^n} \right] &\geq 1-\delta
\\
\trace\left[ \Pi_{\delta}^{(n)}(\rho_{UX_0 X_1 B}|u^n,x_0^n,x_1^n) \omega_{B^n}^{u^n,x_0^n,x_1^n} \right] &\geq 1-\delta
\\
\trace\left[ \Pi_{\delta}^{(n)}(\omega_{UX_0 X_1 B}|u^n,x_0^n,x_1^n)  \right] &\leq 2^{ n(1+2\delta)(H(B|U X_0 X_1)_{\omega} }
\\
\Pi_{\delta}^{(n)}(\omega_{UX_1 B}|u^n,x_1^n)  \omega^{u^n,x_1^n}_{B^n}   \Pi_{\delta}^{(n)}(\omega_{UX_1 B}|u^n,x_1^n) &\leq 2^{ -n(1-2\delta) H(B|U X_1)_{\omega} } \Pi_{\delta}^{(n)}(\omega_{UX_1 B}|u^n,x_1^n)  
\end{align}
imply
 a POVM $\Delta^2_{m_{j}''|u^n,x_1^n}$ such that
\begin{align}
\Pr{ \mathscr{E}_3(j+1) } | { \mathscr{E}_1^c(j)\cap \mathscr{E}_2^c(j) \cap \mathscr{E}_2^c(j-1) } &\leq 2^{ -n( I(X_0;B|U X_1)_\rho -R''-\varepsilon_2) }
\,,
\end{align}
which tends to zero if
\begin{align}
R''< I(X_0;B|U X_1)_\rho -\varepsilon_2 \,.
\label{eq:B3}
\end{align}
This completes the proof of the partial decode-forward lower bound.
\qed %

\section{Proof of Theorem~\ref{Theorem:Hadamard_Relay} (Hadamard Relay Channel)}
\label{Appendix:Hadamard_Relay}
We prove the capacity theorem for the Hadamard relay channel 
$\mathcal{N}_{AD\to BY_1}^{\text{H}}$, where the relay receives a classical observation $Y_1$.
Achievability immediately follows from
Theorem~\ref{theo:PDF}, by taking $U=X_0$ in 
\eqref{eq:low_PDF}.
As explained in Remark~\ref{Remark:Direct_Full_DF}, such a rate can be achieved using a full dicode-forward strategy, where the relay decodes the entire information. That is, we set $R''=0$ and thus $R=R'$ in the proof of Theorem~\ref{theo:PDF} in Appendix~\ref{Appendix:PDF}.

Consider the converse part. 
 Suppose that Alice selects a message $m$ uniformly at random, stores $m$ in a classical register $M$, and then prepares an input state $\rho^{(m)}_{ A^n}$. 
At time $i$, the relay performs an encoding channel, mapping $Y_1^{i-1}$ to $D_i$,
This results in a joint state $\rho_{Y_1^{i-1} D_i A_i   B^{i-1} }^{(m)}$.
Next, the relay receives $Y_{1,i}$ in the state
\begin{align}
\rho_{M B_i  Y_{1,i}  B^{i-1} Y_1^{i-1}  } &=
\frac{1}{2^{nR}}
\sum_{m=1}^{2^{nR}} \ketbra{m}_M \otimes 
(\mathcal{N}^{\text{H}}_{AD\to BY_1}\otimes \mathrm{id}_{  B^{i-1} Y_1^{i-1} })(\rho_{A_i D_i  B^{i-1} Y_1^{i-1} }^{(m)}) 
\,,
\end{align}
for $i=1,2,\ldots,n$. 
Bob receives the channel output systems $B^n$, performs a measurement, and obtains  an estimate  $\widehat{M}$ for Alice's message.

Consider a sequence of codes $(\mathcal{F}_n,\Gamma_n,\Delta_n)$ such that the average probability of error tends to zero.
As usual, we observe that
 Fano's inequality \cite{CoverThomas:06b} implies
$%
H(M|\widehat{M}) \leq n\varepsilon_n
$ and thus %
\begin{align}
nR&
\leq I(M;\widehat{M})+n\varepsilon_n 
\nonumber\\
&\leq I(M;B^n )_{\rho}+n\varepsilon_n
\label{eq:ConvIneq1SC}
\end{align}
based on the data processing inequality for the quantum mutual information, 
where $\varepsilon_n$ tends to zero as $n\rightarrow\infty$. 

Now, we show the following bounds:
\begin{subequations}
\begin{align}
I(M;B^n )_{\rho}\leq 
\sum_{i=1}^n I(M Y_1^{i-1};B_i)_\rho \,,
\label{Equation:Converse_Bound_1_i}
\intertext{and }
I(M;B^n )_{\rho}\leq 
\sum_{i=1}^n I(M;Y_{1,i}|Y_1^{i-1})_\rho \,.
\label{Equation:Converse_Bound_2_i}
\end{align}
\end{subequations}

We begin with the first bound. 
By the chain rule, 
\begin{align}
I(M;B^n )_{\rho}&=\sum_{i=1}^n I(M;B_i|B^{i-1} )_{\rho}
\nonumber\\
&\leq \sum_{i=1}^n I(M B^{i-1};B_i )_{\rho}
\,.
\label{Equation:Converse_Bound_3_i}
\end{align}
Recall that the Hadamard relay channel is degraded, i.e., 
 $\mathcal{N}_{DA\to B Y_1}=
 \mathcal{P}_{Y_1\to BY_1}\circ \mathcal{M}_{DA\to Y_1}$.
 Therefore,
 \begin{align}
  I(M B^{i-1};B_i )_{\rho}\leq I(M Y_1^{i-1};B_i )_{\rho} 
  \label{Equation:Converse_Bound_4_i}
  \end{align}
by the data processing inequality.
The first bound,
\eqref{Equation:Converse_Bound_1_i}, now follows from \eqref{Equation:Converse_Bound_3_i}-\eqref{Equation:Converse_Bound_4_i}.

Similarly, 
\begin{align}
I(M;B^n )_{\rho}&\leq 
I(M;Y_1^n )_{\rho}
\nonumber\\
&=\sum_{i=1}^n I(M;Y_{1,i}|Y_1^{i-1} )_{\rho}
\,,
\label{Equation:Converse_Bound_Fano_DPI_2}
\end{align}
hence the second bound, \eqref{Equation:Converse_Bound_2_i},
holds as well. 

Together with \eqref{eq:ConvIneq1SC}, this implies 
\begin{align}
R-\varepsilon_n\leq 
\min\left\{
\frac{1}{n}\sum_{i=1}^n  I(X_{0,i} X_{1,i};B_i )_{\rho} 
\,,\;
\frac{1}{n}\sum_{i=1}^n  I(X_{0,i};Y_{1,i}|
X_{1,i} )_{\rho} 
\right\}
\label{Equation:Converse_Bound_5_i}
\end{align}
where we have defined 
\begin{align}
X_{0,i}\equiv M
\,,\;
X_{1,i}\equiv Y_1^{i-1}
\,.
\end{align}
Let $J$ to be a uniformly distributed random variable, on $\{1,\ldots, n\}$, which has no correlation with the message.
Then, we may rewrite \eqref{Equation:Converse_Bound_5_i} as
\begin{align}
R-\varepsilon_n\leq 
\min\left\{
 I(X_{0,J} X_{1,J};B_J|J )_{\rho} 
\,,\;
  I(X_{0,J};Y_{1,J}|
X_{1,J} J)_{\rho} 
\right\}
\label{Equation:Converse_Bound_J}
\end{align}
with $\rho_{J X_{0,J} B_J  Y_{1,J}  B^{J-1} X_{1,J}  }=
\frac{1}{n}\sum_{i=1}^n \ketbra{i}_J\otimes \rho_{X_{0,i} B_i  Y_{1,i} B^{i-1} X_{1,i}  }$.
Hence, 
\begin{align}
R-\varepsilon_n\leq 
\min\left\{
 I(X_{0} X_{1};B )_{\rho} 
\,,\;
  I(X_{0};Y_{1}|
X_{1} )_{\rho} 
\right\}
\label{Equation:Converse_Bound_J_1}
\end{align}
for $X_{0}\equiv (J,X_{0,J})$,
$X_1\equiv X_{1,J}$, and thus 
$B\equiv B_J$ and $Y_1\equiv Y_{1,J}$.
This completes the proof of Theorem~\ref{Theorem:Hadamard_Relay}.
\qed %

\section{Proof of Theorem~\ref{theo:MF} (Measure-Forward Strategy)}
\label{Appendix:MF}
Consider a   quantum relay channel $\mathcal{N}_{AD\to BE}$. 
The proof extends the classical scheme of compress-forward coding, along with observations from a previous work by the  author \cite{Pereg:21p}.

We show that for every $\varepsilon_0,\delta_0>0$, there exists a $(2^{n(R-\delta_0)},n,\varepsilon_0)$ code for the quantum relay channel $\mathcal{N}_{AD\to BE}$, provided that $R< \mathsf{R}_{\text{M-F}}(\mathcal{N})$. 
To prove achievability, we extend  the classical block Markov coding to the quantum setting, and then apply the quantum packing lemma and the classical covering lemma. We use the gentle measurement lemma \cite{Winter:99p}, %
which guarantees that multiple decoding measurements can be performed without ``destroying" the output state.

Fix a given input ensemble, $\{ p_{X_0}(x_0) p_{ X_1}(x_1) , \theta_{A}^{x_0}\otimes
\zeta_D^{x_1}\}$, a POVM
$\{\Gamma_{y_1}\}$ on $\mathcal{H}_E$, and a classical channel $p_{Z_1|X_1 Y_1}$. 
 Denote the output states before the relay measurement by
 \begin{align}
\omega^{x_0,x_1}_{BE}&= 
\mathcal{N}_{AD\rightarrow BE}( \theta^{x_0}_{A} \otimes \zeta^{x_1}_D )
\,.
\label{eq:MF_Output}
\end{align}
Upon measurement, the outcome $Y_1$ is distributed according to 
\begin{align}
p_{Y_1|X_0,X_1}(y_1|x_0,x_1)=
\trace( \Gamma_{y_1} \omega_E^{(x_0,x_1)} )\,.
\end{align}
This induces the joint distribution
\begin{align}
p_{X_0 X_1 Y_1 Z_1}(x_0,x_1,y_1,z_1)=
p_{X_0}(x_0)  p_{X_1}(x_1)
p_{Y_1|X_0,X_1}(y_1|x_0,x_1)
p_{Z_1|X_1 Y_1}(z_1|x_1,y_1)
\end{align}
for $(x_0,x_1,y_1,z_1)\in\mathcal{X}_0\times \mathcal{X}_1\times \mathcal{Y}_1\times
\mathcal{Z}_1
$.
Furthemore, consider 
 the average states,
 \begin{align}
  \omega_{AD}^{x_1}&\equiv \left(\sum_{x_0\in\mathcal{X}_0}p_{X_0}(x_0) \theta_A^{x_0}\right)\otimes 
 \zeta_D^{x_1} 
 \,,\;
 \omega_{BE}^{x_1}=
 \mathcal{N}_{AD\to BE}(\omega_{AD}^{x_1})
 \,,
 \\
 \omega_{AD}^{x_0}&\equiv \theta^{x_0}_{A}\otimes 
\left(\sum_{x_1\in\mathcal{X}_1}p_{X_1}(x_1) \zeta_D^{x_1}\right) 
 \,,\;
 \omega_{BE}^{x_0}=
 \mathcal{N}_{AD\to BE}(\omega_{AD}^{x_0})
 \,,
 \\
 \omega_{AD}&\equiv
\left(\sum_{x_0\in\mathcal{X}_0}p_{X_0}(x_0) \theta_A^{x_0}\right)
\otimes 
\left(\sum_{x_1\in\mathcal{X}_1}p_{X_1}(x_1) \zeta_D^{x_1}\right)
\,,\;
\omega_{BE}\equiv 
\mathcal{N}_{AD\to BE}(\omega_{AD})
 \end{align}
 and
 \begin{align}
\omega_{BY_1 Z_1}^{x_0,x_1}=
\sum_{y_1\in\mathcal{Y}_1}
\sum_{z_1\in\mathcal{Z}_1}
p_{Y_1 Z_1|X_0 X_1}(y_1,z_1|x_0,x_1)
\omega_B^{x_0,x_1}\otimes 
\ketbra{y_1}\otimes 
\ketbra{z_1}
\,.
 \end{align}
We use $T$ transmission blocks, where each block consists of $n$ input systems. In particular,  the relay has access to the systems from the previous blocks.
\subsection{Code Construction}
The code construction, encoding and decoding procedures are described below.

\begin{figure*}
\center
\begin{tabular}{l|ccccc}
{\small Block}				
& $1$ & $2$ & $\cdots$ & $T-1$ & $T$ \\
								
\\ \hline &&&&& \\ 
$A$ & $x_0^n(m_1)$	&	$x_0^n(m_2)$ & $\cdots$	& $x_0^n(m_{T-1})$ &  $x_0^n(1)$ 
\\&&&&& \\
$E$	& $\ell_1 \rightarrow$ & $\ell_2 \rightarrow$	& $\cdots$	& $ \ell_{T-1} \rightarrow$ & $\emptyset$
 \\
 & $z_1^n(k_1,\ell_1|1)$ & $z_1^n(k_2,\ell_2|\ell_1)$	& $\cdots$	& $z_1^n(k_{T-1},\ell_{T-1}|\ell_{T-2})$ & $\emptyset$
\\&&&&& \\
$D$	& $x_1^n(1)$ & $x_1^n(\ell_1)$ & $\cdots$	& $x_1^n(\ell_{T-2})$ &  $x_1^n(\ell_{T-1})$
\\&&&&& \\
$B$	& $\emptyset$ & $\hat{\ell}_1$	& $\cdots$	& $\hat{\ell}_{T-2}$ & $\hat{\ell}_{T-1}$
\\
& $\hat{k}_1$ & $\hat{k}_2$	& $\cdots$	& $\hat{k}_{T-1}$ & $\emptyset$
\\
& $\hat{m}_1''$ & $\hat{m}_2''$ & $\cdots$ & $\hat{m}_{T-1}''$ & $\emptyset$
\end{tabular}
\caption[Measure-forward strategy]{Measure-forward strategy. The block index $j\in [1:T]$ is indicated at the top. In the following rows, we have the corresponding elements: 
(1) codewords of Alice;  
(2), (3) measurement representation and
 reconstruction sequence at the relay;
(4) relay codewords;
(5), (6), (7) estimated messages at the destination receiver.
The arrows in the third row indicate that the relay measures and encodes forward.  
}
\label{fig:MF}
\end{figure*}

\subsubsection{Classical Code Construction}
For every $j\in [1:T]$, generate a classical codebook
$\mathscr{B}(j)$ as follows.
Select $ 2^{n R}$ independent sequences
$x_0^n(m_j)$, for 
$m_j\in [1:2^{nR}]$, according to 
$\bar{p}_{X_0^n}$.
Similarly, select $ 2^{n R_1}$ independent sequences
$x_1^n(\ell_{j-1})$, for 
$\ell_{j-1}\in [1:2^{nR_1}]$, according to 
$\bar{p}_{X_1^n}$.
Then, for every given 
$\ell_{j-1}\in [1:2^{nR_1}]$,
select conditionally independent 
$z_1^n(k_j,\ell_j|\ell_{j-1})$,
for 
$k_j\in [1:2^{nR_\text{b}}]$ and 
$\ell_j\in [1:2^{nR_1}]$,
according to 
$ \bar{p}_{Z_1^n|X_1^n}(\cdot|x_1^n)$, conditioned on $x_1^n\equiv x_1^n(\ell_{j-1})$. 

\subsubsection{Encoding}
Set $m_T\equiv 1$.
Given the message sequence 
$(m_j)_{j\in [1:T]}$,
prepare the input state
$\bigotimes_{j=1}^T \rho_{A^n(j)}$, where
\begin{align}
\rho_{A^n(j)}
=
\bigotimes_{i=1}^n
\theta_A^{x_{0,i}}
\,\text{ for $x_{0}^n\equiv 
x_0^n(m_j)$}
\,.
\end{align}
Then, transmit $A^n(j)$ in Block $j$, for $j=1,2,\ldots,T$.

\subsubsection{Relay Encoding}
Set $\ell_0\equiv 1$.
\begin{enumerate}[(i)]
\item
At the end of Block $j$,
perform a measurement on 
$E^n(j)$, using the POVM
$\mathcal{G}_{E\to Y_1}^{\otimes n}$. Denote the measurement outcome by $y_1^n(j)$. 

\item
Find an index pair $(k_j,\ell_j)$ such that
$(y_1^n(j),z_1^n(k_j,\ell_j|\ell_{j-1}),x_1^n(\ell_{j-1}))\in
T_\delta(p_{X_1 Y_1 Z_1})$.
If there is more than one such pair $(k_j,\ell_j)\in [1:2^{nR_\text{b}}]\times [1:2^{nR_1}]$, select the first.
If there is none, set 
$(k_j,\ell_j)=(1,1)$. 
In block $j+1$, prepare the state 
\begin{align}
\rho_{D^n(j+1)}=
\bigotimes_{i=1}^n
\zeta_{D}^{x_{1,i}}
\,,\;
\text{for $x_1^n\equiv x_1^n(\ell_j)$}
\end{align}
using the classical codebook
$\mathscr{B}(j+1)$. Transmit $D^n(j+1)$.
\end{enumerate}
This results in the output state $\bigotimes_{j=1}^T \rho_{B^n(j)}$, where
\begin{align}
&\rho_{B^n(j) }=
\bigotimes_{i=1}^n
\omega_{B}^{x_{0,i},x_{1,i}}
\,,\;
\nonumber\\
&\text{ for 
$x_{0}^n\equiv 
x_0^n(m_j)$ and
$x_1^n\equiv x_1^n(\ell_{j-1})$.}
\end{align}

Bob receives the output  $B^n(j)$ in Block $j$.
\subsubsection{Decoding}
Set $\hat{\ell}_0\equiv 1$.
At the end of Block $j+1$, decode as follows. 
\begin{enumerate}[(i)]
\item
Find an estimate 
$\hat{\ell}_{j}$ by performing a  measurement
$
\{\Delta^1_{\ell_j}\}$,  which will be specified later, on  $B^n(j+1)$.

\item
Find an estimate 
$\hat{k}_j$ by performing a second measurement
$
\{\Delta^2_{k_j|x_1^n}\}$ for 
$x_1^n\equiv x_1^n(\hat{\ell}_{j-1})$,  which will also be specified later, on  $B^n(j)$, for 
$j=1,2,\ldots,T-1$.

\item
Let $\overline{Z}_1^n(j)\equiv 
z_1^n(\hat{k}_j,\hat{\ell}_j|\hat{\ell}_{j-1})$.
Find an estimate 
$\hat{m}_j$ by performing a third measurement
$
\{\Delta^3_{m_j|x_1^n}\}$ for 
$x_1^n\equiv x_1^n(\hat{\ell}_{j-1})$,  which will also be specified later, on  $B^n(j)\overline{Z}_1^n(j)$, for 
$j=1,2,\ldots,T-1$.
\end{enumerate}

\subsection{Error Analysis}
 As we consider Block $j$,
 we may assume without loss of generality that Alice sends the message $M_j=1$, based on the symmetry of the codebook generation.
Consider the following events,
\begin{align}
\mathscr{E}_0(j)&= \{  (X_1^n(L_{j-1}),Z_1^n(k_j,\ell_j|L_{j-1}),Y_1^n(j))\notin
T_\delta(p_{X_1 Y_1 Z_1})
\,\text{ for all $(\ell_j,k_j)\in [1:2^{nR_1}]\times[1:2^{nR_\text{b}}]$}\}
\,,
\\
\mathscr{E}_1(j)&= \{  \widehat{L}_j\neq L_j  \}
\,,
\\
\mathscr{E}_2(j)&= \{  \widehat{K}_j\neq K_j   \}
\,,
\\
\mathscr{E}_3(j)&= \{  \widehat{M}_j\neq 1   \}
\,.
\end{align}
The event $\mathscr{E}_0(j)$ is associated with failure at the relay, and 
$\mathscr{E}_1(j)$, $\mathscr{E}_2(j)$, $\mathscr{E}_3(j)$, with erroneous decoding at the destination receiver.
Let $P_{e}^{(n)}(\mathscr{C},j) $ denote
the probability of error in Block $j$.
By the union of events bound, the expected probability of error  is bounded by
\begin{align}
\mathbb{E} P_{e}^{(nT)}(\mathscr{C}) 
&\leq \sum_{j=1}^T \mathbb{E} P_{e}^{(n)}(\mathscr{C},j)
\,.
\intertext{Furthermore, for every $j\in [1:T]$, }
\mathbb{E} P_{e}^{(n)}(\mathscr{C},j)&\leq 
 \Pr({ \mathscr{E}_0(j-1) })
 +\Pr({ \mathscr{E}_1(j-1) }\cap \mathscr{E}_0^c(j-1))
 +\Pr({ \mathscr{E}_1(j) }\cap \mathscr{E}_0^c(j-1))
\nonumber\\&
+\Pr({ \mathscr{E}_2(j) } \cap { 
\mathscr{E}_0^c(j-1)\cap
\mathscr{E}_1^c(j-1)\cap
\mathscr{E}_1^c(j)})
\nonumber\\&
+\Pr({ \mathscr{E}_3(j) } \cap { 
\mathscr{E}_0^c(j-1)\cap
\mathscr{E}_1^c(j-1)\cap
\mathscr{E}_1^c(j)\cap
\mathscr{E}_2^c(j)
}
)
\label{eq:Pe_MF}
\end{align}
where the conditioning on $M_j=1$ is omitted for convenience of notation. 
We note that the event 
$\mathscr{E}_0^c(j-1)$ is completely classical. 
Then, by the classical covering lemma \cite[Lemma 3.3]{ElGamalKim:11b}, we have
\begin{align}
\Pr({ \mathscr{E}_0(j-1) })\leq 
\exp\left\{ -2^{n(R_1+R_\text{b}- I(Z_1;Y_1|X_1)-\varepsilon_1)} \right\}
\end{align}
which tends to zero, provided that
\begin{align}
R_1+R_\text{b}> I(Z_1;Y_1|X_1)+\varepsilon_1 \,,
\label{Eq:Compression_MF}
\end{align}
where $\varepsilon_1=\varepsilon_1(n)=3\log(\abs{\mathcal{Z}_1})\varepsilon_{\delta_1,\delta}(n)$ tends to zero as $n\to\infty$.

To bound the terms
$\Pr({ \mathscr{E}_1(j-1) }\cap \mathscr{E}_0^c(j-1))
 $ and $\Pr({ \mathscr{E}_1(j) \cap \mathscr{E}_0^c(j-1)})$, we use the quantum packing lemma, Lemma~\ref{lemm:Qpacking}. 
Suppose that the event $\mathscr{E}_0^c(j-1)$ has  occurred. It follows that the relay found $L_{j-1}$ such that the classical sequences are jointly typical, and prepared the input state
according to the classical codeword $X_1^n(L_{j-1})$. 
Observe that based on the quantum typicality properties in Section~\ref{App:Q_Typical} of Appendix~\ref{App:Tools}, 
\begin{align}
\trace\left[ \Pi^{\delta}(\omega_{B}) \omega_{B^n}^{x_1^n} \right] &\geq 1-\delta
\\
\trace\left[ \Pi^{\delta}(\omega_{B}|x_1^n) \omega_{B^n}^{x_1^n} \right] &\geq 1-\delta 
\\
\trace\left[ \Pi^{\delta}(\omega_{B}|x_1^n)  \right] &\leq 2^{ n(H(B| X_1)_{\omega} +2\delta)}  
\\
\Pi^{\delta}(\omega_{B})  \omega_{B}^{\otimes n}   \Pi^{\delta}(\omega_{B}) &\leq 2^{ -n(H(B)_{\omega}-2\delta) } \Pi^{\delta}(\omega_{B})
\end{align}
for all $x_1^n\in T^{(n)}_{\delta_1}( p_{X_1})$ and sufficiently large $n$. Since the codebooks are statistically independent of each other, we have by Lemma~\ref{lemm:Qpacking} that there exists a POVM $\Delta_{\ell_{j}}^1$ such that
$%
\Pr({ \mathscr{E}_1(j) }\cap \mathscr{E}_0^c(j-1))  \leq 2^{ -n( I(X_1;B)_\omega -R_1-\varepsilon_2) } 
$, %
which tends to zero for 
\begin{align}
R_1< I(X_1;B)_\omega -3\varepsilon_2 
\label{eq:R1_MF}
\end{align}
where $\varepsilon_2=\varepsilon_2(\delta)\equiv 4\delta\log(d_B)$.
The same argument holds for $\Pr({ \mathscr{E}_1(j-1) }\cap \mathscr{E}_0^c(j-1))$ as well.

Denote the state of the systems $B^n(j)$ after the measurement above by $\tilde{\rho}_{B^n(j)}$.
By the packing lemma inequality (\ref{eq:QpackB}),  the gentle measurement lemma implies that the post-measurement state is close to the original state in the sense that
\begin{align}
\frac{1}{2}\norm{\tilde{\rho}_{B^n(j)}-\rho_{B^n(j)}}_1 \leq 2^{ -n\frac{1}{2}( I(X_1;B)_\omega -R_1-\varepsilon_2) } \leq 2^{-\varepsilon_2 n}
\end{align}
for sufficiently large $n$ and rates as in (\ref{eq:R1_MF}).
Therefore, the distribution of measurement outcomes when $\tilde{\rho}_{B^n(j)}$ is measured is $2^{-\varepsilon_2 n}$-close to the distribution as if the measurement has never occurred. %

Moving to the third term in the RHS of (\ref{eq:Pe_MF}), suppose that $\mathscr{E}_1^c(j-1)\cap \mathscr{E}_1^c(j)$ occurred as well. Namely, the decoder 
recovered
 $L_{j-1}$ and $L_{j}$ correctly. 
This means that Bob knows 
 $X_1^n(L_{j-1})$.
By the quantum packing lemma, 
 there exists a POVM $\Delta^2_{k_{j}|x_1^n}$ such that
 $
 \Pr({ \mathscr{E}_2(j) }\cap
\mathscr{E}_0^c(j)\cap 
\mathscr{E}_1^c(j-1)\cap \mathscr{E}_1^c(j)) 
\leq
 2^{ -n( I(Z_1;B|X_1)_\omega -R_\text{b}-\varepsilon_2) } $
which tends to zero as $n\rightarrow\infty$, provided that
$R_\text{b}< I(Z_1;B|X_1)_\omega -\varepsilon_2$. Then, let us choose
\begin{align}
R_\text{b}= I(Z_1;B|X_1)_\omega -3\varepsilon_2 \,.
\label{eq:R1p_MF}
\end{align}
By 
the same gentle measurement arguments as before, the post-measurement state is $2^{-\varepsilon_2 n}$-close to the original output state, before the measurement.

We note that given $(x_1,y_1)$, the output state $\omega_{B}^{x_1,y_1}$ has no correlation with $Z_1$. Hence, we may write the requirement in 
\eqref{Eq:Compression_MF} as
\begin{align}
R_1 &> I(Z_1;Y_1 B|X_1)-R_\text{b}+\varepsilon_1
\nonumber\\
 &= I(Z_1;Y_1 |X_1 B)+\varepsilon_1+3\varepsilon_2
\,,
\label{Eq:Compression_MF_1}
\end{align}
where the second line follows from 
\eqref{eq:R1p_MF} and the chain rule.

It remains to consider the last term in the RHS of (\ref{eq:Pe_MF}). If the event $ \mathscr{E}_2^c(j)$ has occurred, then the destination receiver has access to $Z^n(j)\equiv 
Z^n(K_j,L_j|L_{j-1})$. 
By the quantum packing lemma, 
the conditions
\begin{align}
\trace\left[ \Pi^{\delta}(\omega_{BZ_1}|x_1^n) \omega_{B^n Z_1^n}^{x_0^n,x_1^n} \right] &\geq 1-\varepsilon_{4}(\delta)
\\
\trace\left[ \Pi^{\delta}(\omega_{BZ_1}|x_0^n,x_1^n) \omega_{B^n Z_1^n}^{x_0^n,x_1^n} \right] &\geq 1-\varepsilon_{4}(\delta) 
\\
\trace\left[ \Pi^{\delta}(\omega_{BZ_1}|x_0^n)  \right] &\leq 2^{ n(H(BZ_1| X_0)_{\omega} +\varepsilon_{4}(\delta))}  
\\
\Pi^{\delta}(\omega_{BZ_1})  \omega_{B^n Z_1^n}^{x_1^ n}   \Pi^{\delta}(\omega_{BZ_1}|x_1^n) &\leq 2^{ -n(H(B)_{\omega}-\varepsilon_{4}(\delta)) } \Pi^{\delta}(\omega_{BZ_1}|x_1^n)
\end{align}
imply that there exists a POVM $\Delta^3_{m_{j}|x_1^n}$ on
 $B^n(j) Z^n(j)$
 such that
 $
 \Pr({ \mathscr{E}_3(j) }\cap
\mathscr{E}_0^c(j)\cap 
\mathscr{E}_1^c(j-1)\cap \mathscr{E}_1^c(j)\cap \mathscr{E}_2^c(j)) 
\leq
 2^{ -n( I(X_0;B Z_1 |X_1)_\omega -R-\varepsilon_3) } $
which tends to zero as $n\rightarrow\infty$, provided that 
\begin{align}
R< I(X_0;B Z_1|X_1)_\omega -\varepsilon_3 
\label{eq:B2_Rp}
\end{align}
where $\varepsilon_2=\varepsilon_2(\delta)\equiv 4\delta\log(\abs{\mathcal{Z}_1}d_B)$.
By eliminating $R_1$, we obtain 
 the measure-forward lower bound.
\qed %

\section{Proof of Theorem~\ref{theo:AF} (Assist-Forward Strategy)}
\label{Appendix:AF}
Consider a   quantum relay channel with ORC, $B=(B_1,B_2)$, where
the channel map $\mathcal{N}_{AD\to B_1 B_2 E}$ has the following form:
$%
\mathcal{N}_{AD\to B_1 B_2 E}=
\mathcal{M}_{A\to B_1 E}\otimes
\mathcal{P}_{D\to B_2} 
$ (as in \eqref{Equation:Ortho_Receiver}). %
We introduce a coding scheme where the transmitter (Alice) generates entanglement assistance between the relay and the destination receiver (Bob). This enables entanglement-assisted communication from the relay to Bob in the subsequent block. 
Our proof combines block Markov coding with various techniques in quantum information theory, by 
Winter \cite{Winter:99p}, Dupuis et al. \cite{DupuisHaydenLi:10p}, and Shor \cite{Shor:04p},
on 
constant-composition codes, broadcast subspace transmission, and rate-limited entanglement assistance, respectively.

We show that for every $\varepsilon_0,\delta_0>0$, there exists a $(2^{n(R-\delta_0)},n,\varepsilon_0)$ code for the quantum relay channel $\mathcal{N}_{DA\to B_1 B_2E}$, provided that $R< \mathsf{R}_{\text{A-F}}(\mathcal{N})$. 
Fix a pair of types, $p_{X_1}$ and $ p_{X_2}$, and input ensembles, $\{ p_{X_1}(x_1) p_{X_2}(x_2) , \ket{\theta_{G_0 G_1 A}^{x_1}}\otimes
\ket{\zeta_{G_2 D}^{x_2}}\}$. Denote the output states by 
\begin{align}
\theta_{G_0 G_1 B_1 E}^{x_1}&\equiv 
(\mathrm{id}_{G_0 G_1}\otimes \mathcal{M}_{A\to B_1 E})(\ketbra{\theta_{G_0 G_1 A}^{x_1}}) \,,
\\
\zeta_{G_2 B_2}^{x_2}&\equiv 
(\mathrm{id}_{G_2}\otimes \mathcal{P}_{D\to B_2})(\ketbra{\zeta_{G_2 D}^{x_2}}) \,,
\end{align}
 for $(x_1,x_2)\in\mathcal{X}_1\times\mathcal{X}_2$,
and
the average states:
\begin{align}
\theta_{G_0 G_1 A}&\equiv
 \sum_{x_1\in\mathcal{X}_1} p_{X_1}(x_1) 
\ketbra{\theta_{G_0 G_1 A}^{x_1}} \,,\;
\zeta_{G_2 D}\equiv
 \sum_{x_2\in\mathcal{X}_2} p_{X_2}(x_2) 
\ketbra{\zeta_{G_2 D}^{x_2}} \,,
\\
\theta_{G_0 G_1 B_1 E}&\equiv
(\mathrm{id}_{G_0 G_1}\otimes \mathcal{M}_{A\to B_1 E})(\ketbra{\theta_{G_0 G_1 A}}) \,,\;
\zeta_{G_2 B_2}\equiv
(\mathrm{id}_{G_2}\otimes \mathcal{P}_{D\to B_2})(\ketbra{\zeta_{G_2 D}}) \,.
\end{align}

Recall that the relay encodes in a strictly-causal manner. Specifically, the relay  has access to the sequence of previously received systems, $E_{i-1},\bar{E}^{i-2}$, from the past.
We use $T$ transmission blocks, where each block consists of $n$ input systems. In particular,  the relay has access to the systems from the previous blocks.
In effect, the $j^{\text{th}}$ transmission block of the relay encodes  the message $m_{j-1}\in [1:2^{nR}]$ from the previous block.

\subsection{Code Construction}

The code construction, encoding and decoding procedures are described below.

\subsubsection{Classical Codebook Construction}
For every $j\in [1:T]$, generate a classical codebook
$\mathscr{B}_1(j)$ as follows.
Select $ 2^{n R}$ independent sequences
$x_1^n(m_{j})$,
for 
$m_{j}\in [1:2^{nR}]$, each drawn uniformly at random from the  type class $\mathcal{T}^{(n)}(p_{X_1})$.
Assume without loss of generality that 
$\mathcal{X}_1=\{\alpha_1,\alpha_2,\ldots,\alpha_\abs{\mathcal{X}_1}\}$.
Let $\pi_{m_j}$ be a permutation that rearranges the codeword $x_1^n(m_{j})$ in a lexicographic order, i.e.
\begin{align}
\pi_{m_j} [x_1^n(m_{j}))]=
\big( &\alpha_1,\ldots,\alpha_1 ,\\
&\alpha_2,\ldots,\alpha_2 ,\\
&\ldots\\
&\alpha_\abs{\mathcal{X}_1},\ldots,\alpha_\abs{\mathcal{X}_1}\big) \,.
\end{align}
The number of appearances of  each letter $\alpha\in\mathcal{X}_1$ is
\begin{align}
n_\alpha \equiv np_{X_1}(\alpha)
\,,
\end{align}
since all codewords are of the same type, $p_{X_1}$.
Applying the permutation operator $\pi_{m_j}$ on the state
\begin{align}
 \theta_{A^n}^{x_1^n} =\bigotimes_{i=1}^n \theta_A^{x_{1,i}} 
\,,\,\text{ for $x_{1}^n\equiv 
x_1^n(m_j)$,}
\end{align}
yields
\begin{align}
\varphi_{A^n(j)}
&= \pi_{m_j} \left[ \theta_{A^n}^{x_1^n} \right] 
\nonumber\\
&=
\bigotimes_{\alpha\in\mathcal{X}_1}
\left(\theta_A^{\alpha} \right)^{\otimes n_\alpha}
\end{align}
for all $m_j\in [1:2^{nR}]$.
This follows the approach by Wilde \cite[Sec. 22.5.1]{Wilde:17b}.
 
\subsubsection{Encoding}
Set $m_1=m_2=m_T\equiv 1$. Hence, we encode in an average rate of $\left( \frac{T-3}{T} \right)R$, which tends to $R$ as $T\to \infty$.
For every $\alpha\in\mathcal{X}_1$, Alice encodes a sub-block of length $n_\alpha$ as follows. She prepares $n_\alpha Q_\alpha$ maximally entangled qubit pairs, i.e., $n_\alpha Q_\alpha$ local EPR pairs. Denote the overall state by 
$\ket{\Psi^{(n_\alpha)}_{L_0 L_1}}$.
She applies a broadcast encoder $\mathcal{F}^{(\alpha)}_{L_0 L_1\to A^n}$, which will be prescribed later. 

We will choose an encoder $\mathcal{F}^{(\alpha)}_{L_0 L_1\to A^n}$ that distributes the entangled systems $L_0$ and $L_1$ to Bob and the relay, respectively. 
Based on the quantum broadcast results by Dupuis et al. \cite{DupuisHaydenLi:10p}, this can be obtained provided that the qubit rate for $L_0$ and $L_1$ is bounded by 
$I(G_0\rangle B_1)_{\theta^\alpha}$ and 
$I(G_1\rangle E)_{\theta^\alpha}$, respectively (see Theorem~4 therein).
Furthermore,  the input state is $\varepsilon_{1}$-close to $(\theta_A^\alpha)^{\otimes n_\alpha}$ in trace distance, for sufficiently large $n$ and $\varepsilon_{1}\equiv 
\varepsilon_{1}(n,\delta)$ that tends to zero as $n\to\infty$ and $\delta\to 0$ \cite[Sec. VI]{DupuisHaydenLi:10p}.
Then, Alice applies the inverse permutation,
$\pi_{m_j}^{-1}$, and transmits $A^n(j)$ over the broadcast channel 
$\mathcal{M}_{A\to B_1 E}^{\otimes n}$
in Block $j$, for $j=1,2,\ldots,T$.
The channel input is thus $\varepsilon_{1}$-close to $\theta_{A^n}^{x_1^n(m_j)}$.

\subsubsection{Relay Encoding}
Set $\tilde{m}_1=\tilde{m}_2=\tilde{m}_T\equiv 1$.
For 
$j=2,\ldots,T-1$:
\begin{enumerate}[(i)]
\item
At the end of Block $j$,
find an estimate $\tilde{m}_j$ by performing a measurement $\{\Gamma_{m_j}\}_{m_j\in [1:2^{nR}]}$, which will be specified later, on the received systems
$E^n(j)$.

\item
In Block $j+1$,
send the message $m_j$ to Bob, using the entanglement assistance from Block $j-1$.

\item
At the end of Block $j+1$,
apply the permutation operator $\pi_{m_j}$ on $E^n(j+1)$.
For every $\alpha\in\mathcal{X}_1$, decode the entanglement resource $L_1$, using a decoding map $\mathcal{G}^{(\alpha)}_{E^{n_\alpha}\to \hat{L}_1}$, which will also be prescribed later.

\end{enumerate}

Bob receives the $T$ output pairs, $(B_1^n(j),B_2^n(j))$, for 
$j\in [1:T]$. %
\subsubsection{Decoding}
Set $\hat{m}_0=\hat{m}_1=\hat{m}_2\equiv 1$. 
\begin{enumerate}[(i)]
\item
Perform a decoding measurement on the second output $B_2^n(j+1)$ and the previously-decoded entanglement resource $L_0(j-1)$ in order to decode $\hat{m}_{j}$, for $j= 2,\ldots,T-1$.

\item
Apply the permutation 
$\pi_{\hat{m}_{j}}$ on $B_1^n(j)$.
For every $\alpha\in\mathcal{X}_1$, decode the entanglement resource $L_0(j)$, using a decoding map $\mathcal{D}^{(\alpha)}_{B_1^{n_\alpha}\to \hat{L}_0}$, which will  be prescribed later, $j= 1,\ldots,T$. 

\end{enumerate}

\subsection{Error Analysis}
Consider the relay encoding.
First, we use a constant-composition version of  the HSW Theorem \cite{Holevo:98p,SchumacherWestmoreland:97p}. 
Based on \cite[Th. 10]{Winter:99p}, 
there exists a measurement $\{\Gamma_{m_j}\}_{m_j\in [1:2^{nR}]}$ on  $E^n(j)$ such that the relay can recover $m_j$ with vanishing probability of error, provided that
\begin{align}
R< I(X_1;E)_\theta-\varepsilon_{2}(n,\delta)
\end{align}
for sufficiently large $n$, where 
$\varepsilon_{2}(n,\delta)$ tends to zero as $n\to\infty$ and $\delta\to 0$.
The gentle measurement lemma 
{\cite{Winter:99p,OgawaNagaoka:07p}}, Lemma~\ref{lemm:gentleM},
 implies that the post-measurement state is $\varepsilon_3$-close to the original state before the measurement, where $\varepsilon_{3}\equiv 
\varepsilon_{3}(n,\delta)$ tends to zero as well.

Having recovered the message, both the relay and Bob apply the permutation operator $\pi_{m_j}$, and the use broadcast decoders, $\mathcal{G}^{(\alpha)}_{E^{n_\alpha}\to \hat{L}_1}$ and 
$\mathcal{D}^{(\alpha)}_{B_1^{n_\alpha}\to \hat{L}_0}$, respectively, for $\alpha\in\mathcal{X}_1$, in order to decode the entanglement resources. 
We now apply the results by Dupuis et al. \cite{DupuisHaydenLi:10p} on quantum subspace transmission via a broadcast channel. Here, our broadcast channel is the channel $\mathcal{M}_{A\to B_1 E}$ from Alice to the relay and Bob's first component, $B_1$.
By \cite[Th. 4]{DupuisHaydenLi:10p}, the relay and destination receiver can recover the quantum resources with vanishing trace-distance error and entanglement rate $Q$, if
\begin{align}
Q &< \frac{1}{n} \sum_{\alpha\in\mathcal{X}_1} n_\alpha
I(G_0 \rangle B_1 | X_1=\alpha)_\theta 
-\varepsilon_{4}(n,\delta)
\,,
\intertext{and}
Q &< \frac{1}{n} \sum_{\alpha\in\mathcal{X}_1} n_\alpha
I(G_1 \rangle E | X_1=\alpha)_\theta -\varepsilon_{4}(n,\delta)
\,,
\end{align}
for sufficiently large $n$, where $\varepsilon_{4}(n,\delta)$ tends to zero as $n\to\infty$ and $\delta\to 0$.
We can also write this requirement as
\begin{align}
Q &< \min \left\{
\sum_{\alpha\in\mathcal{X}_1} p_X(\alpha)
I(G_0 \rangle B_1 | X_1=\alpha)_\theta
\,,\;
\sum_{\alpha\in\mathcal{X}_1} p_X(\alpha)
I(G_1 \rangle E | X_1=\alpha)_\theta
\right\}-\varepsilon_{4}(n,\delta)
\nonumber\\
&= \min \left\{
I(G_0 \rangle B_1 X_1 )_\theta
\,,\;
I(G_1 \rangle E X_1 )_\theta
\right\}-\varepsilon_{4}(n,\delta)
\,.
\end{align}

Then, the relay can communicate to Bob over the channel 
$\mathcal{P}_{D\to B_2}$ with entanglement assistance, limited to the rate $Q$.
Based on the Shor's result \cite{Shor:04p} \cite[Th. 25.2.6]{Wilde:17b} on rate-limited entanglement assistance, the relay can thus send the message $m_j$ to Bob with vanishing probability of error, provided that
\begin{align}
R &< I(X_2 G_2;B_2)_\zeta-\varepsilon_{5}(n,\delta) \,,
\\
R &< I(X_2;B_2)_\zeta+ I(G_2\rangle B_2 X_2)_\zeta +Q-\varepsilon_{5}(n,\delta)
\end{align}
where %
$\varepsilon_{5}(n,\delta)$ tends to zero.
This completes the proof of the assist-forward lower bound.
\qed %

\section{Depolarizing Relay Channel}
\label{Appendix:Depolarizing}
Consider the quantum relay channel in Example~\ref{Example:Depolarizing}.
According to the measure-forward bound in 
Theorem~\ref{theo:MF},
\begin{align}
C(\mathcal{N})\geq
\mathsf{R}_{\text{M-F}}(\mathcal{N}) \,.
\end{align}
Set 
$X_0$ and $X_1$ according to
$\text{Bernoulli}\left( \frac{1}{2} \right)$ over the computational basis ensembles.
This results in the following output states,
\begin{align}
\omega^{(x_0=0)}_{B_1 E}&= 
\frac{1}{2}\Big[
\ketbra{0}\otimes \theta_0+
\ketbra{1}\otimes
\mathsf{X} \theta_0 \mathsf{X}
\Big]
\,,\;
\omega^{(x_1=0)}_{B_2}= 
(1-q)\ketbra{0}+q\ketbra{1}
\,,
\\
\omega^{(x_0=1)}_{B_1 E}&= 
\frac{1}{2}\Big[
\ketbra{1}\otimes \theta_0+
\ketbra{0}\otimes
\mathsf{X} \theta_0 \mathsf{X}
\Big]
\,,\;
\omega^{(x_1=0)}_{B_2}= 
(1-q)\ketbra{0}+q\ketbra{1}
\,.
\label{eq:MF_Output_1_Depolarizing}
\end{align}

Suppose that the relay performs a measurement $\{\Gamma_{y_1}\}$ on $E$ in the computational basis. This results in a measurement outcome $Y_1%
$, where
\begin{align}
\omega^{(x_0=0)}_{B_1 Y_1}&= 
\frac{1}{2}\Big[
\ketbra{0}\otimes \left( (1-p)\ketbra{0}_{Y_1}+p\ketbra{1}_{Y_1}  \right)+
\ketbra{1}\otimes
\left( (1-p)\ketbra{1}_{Y_1}+p\ketbra{0}_{Y_1}  \right)
\Big]
\nonumber\\
&=
\frac{1}{2}\Big[
\left( (1-p)\ketbra{0}+p\ketbra{1} \right)\otimes  \ketbra{0}_{Y_1}
+
\left( (1-p)\ketbra{1}+p\ketbra{0} \right)\otimes  \ketbra{1}_{Y_1}
\Big]
\,,
\\
\omega^{(x_0=1)}_{B_1 Y_1}&= 
\frac{1}{2}\Big[
\ketbra{1}\otimes \left( (1-p)\ketbra{0}_{Y_1}+p\ketbra{1}_{Y_1}  \right)+
\ketbra{0}\otimes
\left( (1-p)\ketbra{1}_{Y_1}+p\ketbra{0}_{Y_1}  \right)
\Big]
\nonumber\\
&=
\frac{1}{2}\Big[
\left( (1-p)\ketbra{1}+p\ketbra{0} \right)\otimes  \ketbra{0}_{Y_1}
+
\left( p\ketbra{1}+(1-p)\ketbra{0} \right)\otimes  \ketbra{1}_{Y_1}
\Big]
\,,
\label{eq:MF_Output_2_Depolarizing}
\end{align}

Now, let
\begin{align}
Z_1=Y_1+V_1 \mod 2
\end{align}
with $V_1\sim\text{Bernoulli}\left( \alpha \right)$, where the parameter $\alpha$ will be chosen later. Then, similarly, 
\begin{align}
\omega^{(x_0=0)}_{B_1 Z_1}&= 
\frac{1}{2}\Big[
\left( (1-\alpha*p)\ketbra{0}+(\alpha*p)\ketbra{1} \right)\otimes  \ketbra{0}_{Z_1}
+
\left( (1-\alpha*p)\ketbra{1}+(\alpha*p)\ketbra{0} \right)\otimes  \ketbra{1}_{Z_1}
\Big]
\,,
\\
\omega^{(x_0=1)}_{B_1 Z_1}&= 
\frac{1}{2}\Big[
\left( (1-\alpha*p)\ketbra{1}+(\alpha*p)\ketbra{0} \right)\otimes  \ketbra{0}_{Z_1}
+
\left( (\alpha*p)\ketbra{1}+(1-\alpha*p)\ketbra{0} \right)\otimes  \ketbra{1}_{Z_1}
\Big]
\,.
\label{eq:MF_Output_3_Depolarizing}
\end{align}
Hence,
\begin{align}
I(X_0;Z_1 B_1 B_2|X_1)_\omega
&=
I(X_0;Z_1 B_1)_\omega
\nonumber\\
&=I(X_0; B_1|Z_1)_\omega
\nonumber\\
&=1-h(\alpha*p)_\omega
\end{align}
where the first equality holds since
$I(X_0; B_2|X_1 Z_1 B_1)_\omega=0$, and the second since $I(X_0; Z_1)_\omega=0$.
We deduce that 
\begin{align}
C(\mathcal{N})\geq 1-h(\alpha*p)_\omega
\end{align}
provided that the maximization constraint is satisfied. 

The maximization constraint requires that we choose $\alpha$ such that
\begin{align}
I(Z_1;Y_1| B_1)_\omega \leq I(X_1;B_2)_\omega
\,.
\label{Equation:I_Z1_Y1_B_Depolarizing}
\end{align}
On the right-hand side, we have
\begin{align}
I(X_1;B_2)_\omega=1-h\left( \frac{q}{2} \right)
\,.
\end{align}
On the left-hand side,
\begin{align}
I(Z_1;Y_1| B_1)_\omega&=I(Z_1;Y_1)_\omega+I(Z_1; B_1 |Y_1)_\omega
-I(Z_1; B_1)_\omega
\nonumber\\
&=I(Z_1;Y_1  )_\omega
\nonumber\\
&=1-h(\alpha)
\,,
\end{align}
where the first equality holds by the chain rule, and the second  since
$I(Z_1; B_1 |Y_1)_\omega
=I(Z_1; B_1)_\omega=0$.
Hence, the constraint in  \eqref{Equation:I_Z1_Y1_B_Depolarizing} is met for $\alpha=\frac{q}{2}$. This completes the derivation.

}
\end{appendices}

\bibliography{references}%

\end{document}